\documentclass[%
 reprint,
preprintnumbers,
 amsmath,amssymb,
 aps,
]{revtex4-2}

\usepackage[english]{babel}
\usepackage{graphicx}  
\usepackage{dcolumn}   
\usepackage{bm}        
\usepackage{grffile}
\usepackage{xspace}
\usepackage{ifthen}
\usepackage{xifthen}
\usepackage{afterpage}
\usepackage{delimset}
\usepackage{stmaryrd}
\usepackage{xcolor}
\usepackage{comment}

\newcommand{\fvec}  [1]    {\boldsymbol{#1}}

\newcommand{\diff}         {\operatorname{d}\!}

\newcommand{\prob}  [2][]  {%
  \ensuremath{%
    \mathbb{P}\ifthenelse{\isempty{#1}}%
      {\left({#2}\right)}%
      {\left({#2}\,\middle|\,{#1}\right)}%
  }\xspace%
}

\newcommand{\cprob}     [2]{\mathbb{P}\left({#1}\,\middle|\,{#2}\right)}

\newcommand{\Afrag}        {{A_{\rm f}}}
\newcommand{\Zfrag}        {{Z_{\rm f}}}
\newcommand{\Nfrag}        {{N_{\rm f}}}
\newcommand{\Xfrag}        {{X_{\rm f}}}

\newcommand{\ZR}           {{Z_{\rm R}}}
\newcommand{\NR}           {{N_{\rm R}}}
\newcommand{\opAR}         {{\op{A}_{\rm R}}}
\newcommand{\opZR}         {{\op{Z}_{\rm R}}}
\newcommand{\opNR}         {{\op{N}_{\rm R}}}

\newcommand{\ProjZ}        {\op{P}_{\rm p}}
\newcommand{\ProjZR}       {\op{P}_{\rm p}^{\rm (R)}}
\newcommand{\ProjN}        {\op{P}_{\rm n}}
\newcommand{\ProjNR}       {\op{P}_{\rm n}^{\rm (R)}}

\newcommand{\zneck}        {z_{\rm N}}
\newcommand{\aneck}        {a_{\rm N}}
\newcommand{\qneck}        {q_{\rm N}}
\newcommand{\qnecksciss}   {q_{\rm N}^{\rm sciss}}
\newcommand{\Qneck}        {\op{Q}_{\rm N}}

\newcommand{\op}    [1]    {\ensuremath{\hat{#1}}}

\newcommand{\ket}   [1]    {\ensuremath{\left|{#1}\right\rangle}\xspace}
\newcommand{\braket}[3][]  {%
  \ensuremath{%
    \ifthenelse{\isempty{#1}}{%
      \left\langle{#2}\middle|{#3}\right\rangle%
    }{%
      \delimtriple<\vert\vert>!{#2}{#1}{#3}%
    }%
  }\xspace%
}
\newcommand{\qavg}  [2][]  {\ensuremath{\left\langle\xspace{#2}\xspace\right\rangle_{#1}}}

\newcommand{\Hgcm}         {\mathcal{H}_{\rm coll}}
\newcommand{\GCMExtraEneInit} {\bar{E}}
\newcommand{\GCMSigInit} {\sigma}

\newcommand{\idHFB} [1]    {\phi({#1})}

\newcommand{\idGCM} [1]    {\Phi({#1})}
\newcommand{\idPV}         {0}

\newcommand{\ladder}[4]    {%
  \def\tempid{#1}%
  \def\tempac{#2}%
  \ifx\tempid\empty%
    {#3}_{#4}\ifx#2c^{\dagger}\fi%
  \else%
    {#3}^{(#1)\ifx#2c{\dagger}\fi}_{#4}%
  \fi%
}
\newcommand{\ladderc}[4]    {%
  \def\tempid{#1}%
  \def\tempac{#2}%
  \ifx\tempid\empty%
    {#3}({#4})\ifx#2c^{\dagger}\fi%
  \else%
    {#3}^{(#1)\ifx#2c{\dagger}\fi}({#4})%
  \fi%
}
\newcommand{\idPw}         {a}
\newcommand{\aPw}   [2][]  {\ladder{#1}{a}{\hat{\idPw}}{#2}}
\newcommand{\cPw}   [2][]  {\ladder{#1}{c}{\hat{\idPw}}{#2}}
\newcommand{\aPr}   [2][]  {\ladderc{#1}{a}{\hat{\idPw}}{#2}}
\newcommand{\cPr}   [2][]  {\ladderc{#1}{c}{\hat{\idPw}}{#2}}

\newcommand{\idQP}         {\beta}
\newcommand{\aQP}   [2][]  {\ladder{#1}{a}{\hat{\idQP}}{#2}}

\newcommand{\figref}  [1]{Fig.~\ref{#1}}

\newcommand{\ssecref} [1]{Sec.~\ref{#1}}

\DeclareMathOperator\erf{erf}

\begin{document}

\title{Microscopic Calculation of Fission Product Yields with Particle Number Projection}
\author{Marc Verriere}%
 \email{verriere1@llnl.gov}
 \affiliation{%
 Nuclear and Chemical Sciences Division, Lawrence Livermore National Laboratory, Livermore, California 94551, USA
 }
 
\author{David Regnier}
 \affiliation{CEA, DAM, DIF, 91297 Arpajon, France}
 \affiliation{Universit\'e Paris-Saclay, CEA, LMCE, 91680 Bruy\`eres-le-Ch\^atel, France}
 
\author{Nicolas Schunck}
 \affiliation{%
 Nuclear and Chemical Sciences Division, Lawrence Livermore National Laboratory, Livermore, California 94551, USA
 }
 
\date{\today}
\preprint{LLNL-JRNL-818368}

\begin{abstract}
Fission fragments' charge and mass distribution is an important input to applications ranging from basic science to energy production or nuclear non-proliferation. In simulations of nucleosynthesis or calculations of superheavy elements, these quantities must be computed from models, as they are needed in nuclei where no experimental information is available. Until now, standard techniques to estimate these distributions were not capable of accounting for fine-structure effects, such as the odd-even staggering of the charge distributions. In this work, we combine a fully-microscopic collective model of fission dynamics with a recent extension of the particle number projection formalism to provide the highest-fidelity prediction of the primary fission fragment distributions for the neutron-induced fission of $^{235}$U and $^{239}$Pu. We show that particle number projection is an essential ingredient to reproduce odd-even staggering in the charge yields and benchmark the performance of various empirical probability laws that could simulate its effect.
This new approach also enables for the first time the realistic determination of two-dimensional isotopic yields within nuclear density functional theory.
\end{abstract}


\maketitle


\section{Introduction}
A predictive theory of nuclear fission has been a longstanding challenge of nuclear science that has gained renewed interest in recent years \cite{bender2020}. While fission is a fascinating problem on its own, as it involves the large-amplitude collective dynamics of a strongly-interacting quantum many-body system, it also plays an important role in both fundamental science and technological applications. For example, fission is a primary decay mechanism of superheavy elements \cite{baran2015} and plays a crucial role in the rapid neutron capture process at the origin of heavy elements in the universe \cite{mumpower2020primary}. Progress in these disciplines requires accurate and precise fission data, such as the distribution and full characteristics (charge, mass, excitation energy, spin, level density, etc.) of the fragments formed during the process. However, the ensemble of all these fissioning nuclei covers a vast area of the nuclear chart. While precise data are available on experimentally accessible nuclei, information on very short-lived, neutron-rich systems out of reach of experimental facilities must come from theoretical predictions.  

There are many approaches geared toward describing low-energy neutron-induced fission reactions~\cite{schunck2016microscopic,madland1988new,berger1991time,tanimura2017microscopic,bulgac2016induced,moller2016nuclear,goutte2005microscopic,lemaitre2019fully,schmidt2016general}. Among them, collective models have been particularly successful in predicting fission fragment distributions~\cite{moller2001,goutte2005microscopic,moller2012,moller2015,regnier2016fission,tao2017,regnier2019,zhao2019a,mumpower2020}. These models are based on the identification of a few collective variables driving the fission process, the calculation of a potential energy surface in the resulting collective space, and the explicit time-dependent simulation of collective motion on top of these surfaces \cite{schunck2016microscopic}. Basic fission fragment properties such as proton- or neutron-number are mapped to the fissioning nucleus' characteristics, such as its deformation. Until now, particle number estimates were obtained by simply integrating the density of particles in the prefragments. This local averaging made it impossible to predict fine structure effects such as the odd-even staggering in the fission fragments charge distributions. 

In the seminal work of Refs.~\cite{simenel2010particle,scamps2015superfluid}, the authors introduced a new method based on particle-number projection techniques to predict particle transfer in heavy-ion reactions in the context of time-dependent density functional theory. In \cite{verriere2019number}, this method was applied to calculate the dispersion in particle number for the most probable scission configuration of $^{239}$Pu(n,f) fission and showed that odd-even staggering naturally emerged. By further combining particle-number projection in the fission fragments with a strongly-damped random walk on semi-classical potential energy surfaces, the authors of \cite{verriere2020improvements} showed that it is possible to predict odd-even staggering but also the charge polarization of the fission fragments distribution.

This paper aims to combine particle-number projection in fission fragments with a  quantum-mechanical theory of large-amplitude collective dynamics to predict the uncorrelated mass and charge fission fragment distributions before prompt emission. We investigate the role of projection for reproducing odd-even staggering effects in fragment distributions and discuss how various phenomenological probability distributions could approximate exact results. 
We also present the first two-dimensional isotopic yields predicted with such a microscopic approach and their evolution as a function of excitation energy.

In Section \ref{sec:theo}, we describe in detail our theoretical framework. It includes a short reminder on the time-dependent generator coordinate method under the Gaussian overlap approximation with Hartree-Fock-Bogoliubov generator states, a comprehensive presentation of the method used to extract fission yields from the time evolution of a collective wave packet, and a discussion of the various methods to estimate particle number dispersion in prefragments. Section \ref{sec:res} contains the results of our calculations for the two important cases of $^{235}$U(n,f) and $^{239}$Pu(n,f) low-energy fission, focusing on the impact of particle number projection and the evolution of the yields as a function of excitation energy.

\section{Theoretical framework}
\label{sec:theo}
Our goal is to predict the initial, or primary, fission fragment mass and charge distributions before prompt emission of neutrons and gammas.
These quantities are determined by populating scission configurations by solving a collective Schr\"odinger-like equation in the collective space spanning nuclear deformations.
This method, presented in \ssecref{sec:theo:dynamic}, relies on (i) defining a set of collective variables and (ii) a basis of many-body generator states calculated with the constrained Hartree-Fock-Bogoliubov method.
They are used both in the definition of the potential energy surface and of the collective inertia.
In \ssecref{sec:theo:proj}, we present how fission yields can be extracted by combining the population of scission configurations and an estimate of particle number distributions at each configuration. 

\subsection{Collective Dynamics}
\label{sec:theo:dynamic}

Our model of fission dynamics is based on the time-dependent generator coordinate method (TDGCM) under the Gaussian overlap approximation (GOA) with Hartree-Fock-Bogoliubov (HFB) generator states~\cite{verriere2020timedependent}. 

\subsubsection{The Time-Dependent Generator Coordinate Method}
\label{sec:theo:tdgcm}

The central assumption of the TDGCM is the possibility to decompose the many-body wave function $\ket{\idGCM{t}}$, at each time $t$, as a superposition of many-body states $\ket{\idHFB{\fvec{q}}}$. It reads
\begin{equation}
    \ket{\idGCM{t}} = \int\diff{\fvec{q}} \ket{\idHFB{\fvec{q}}}f(\fvec{q}, t)\ ,
\end{equation}
where $\fvec{q}$ labels the collective degrees of freedom associated, in our case, with average deformations of the fissioning system.
The complex-valued weights $f(\fvec{q}, t)$ can be explicitly determined by solving the non-local Hill-Wheeler-Griffin (HWG) equation~\cite{verriere2017fission,verriere2020timedependent}.
However, such an approach is extremely time-consuming.
Consequently, we adopt the Gaussian Overlap Approximation (GOA) to transform the HWG equation into a local, Schr\"odinger-like equation,
\begin{equation}\label{eq:theo:tdgcm:tdgcmgoa}
    i\hbar\frac{\partial}{\partial t}g(\fvec{q}, t) = \Hgcm(\fvec{q})g(\fvec{q}, t)\ ,
\end{equation} 
where $g(\fvec{q}, t)$ is an unknown complex-valued function that encodes the collective dynamics. The collective Hamiltonian $\Hgcm(\fvec{q})$ is defined as
\begin{equation}
    \Hgcm(\fvec{q}) = -\frac{\hbar^2}{2\sqrt{\gamma(\fvec{q})}}\nabla_{\fvec{q}}\sqrt{\gamma(\fvec{q})}B(\fvec{q})\nabla_{\fvec{q}} + V(\fvec{q})\ .
\end{equation}
In this expression, $\gamma(\fvec{q})$ is the metric of the collective space, $B(\fvec{q})$ is the inertia tensor and $V(\fvec{q}) = E(\fvec{q})-\varepsilon(\fvec{q})$, where $E(\fvec{q})$ is the HFB energy of the fissioning system at point $\fvec{q}$ and $\varepsilon(\fvec{q})$ is a sum of zero-point-energy corrections. All these quantities are extracted from the states $\ket{\idHFB{\fvec{q}}}$ and a nuclear energy functional  that encodes the nucleon-nucleon interactions; see \cite{ring2004,krappe2012,schunck2016microscopic} for additional details. 

To ensure that we can describe asymmetric fission, where the existence of light and heavy fragments can be mapped to parity-breaking shapes of the fissioning nucleus at scission, we will work in a two-dimensional collective space spanned by the generic vector $\fvec{q} = (q_0, q_1)$, where the first degree of freedom, $q_0$, is the average of the quadrupole moment operator $\op{Q}_{20}$ on the generator state $\ket{\idHFB{\fvec{q}}}$, and the second one, $q_1$, is the average of the octupole moment operator $\op{Q}_{30}$ on that same state,
\begin{equation}\label{eq:constraints}
    q_0 = \frac{\braket[\op{Q}_{20}]{\idHFB{\fvec{q}}}{\idHFB{\fvec{q}}}}{\braket{\idHFB{\fvec{q}}}{\idHFB{\fvec{q}}}}
    \qquad
    q_1 = \frac{\braket[\op{Q}_{30}]{\idHFB{\fvec{q}}}{\idHFB{\fvec{q}}}}{\braket{\idHFB{\fvec{q}}}{\idHFB{\fvec{q}}}}\ .
\end{equation}
In practical calculations, an additional constraint on the dipole moment operator $\op{Q}_{10}$ must be added to fix the position of the center-of-mass of the nucleus.

The initial collective wave-packet $g(\fvec{q},t=0)$ is constructed from the eigenstates $g_k^{\text{extra}}(\fvec{q})$ with eigenvalues $E_k$ of the static GCM+GOA equation in an extrapolated first potential well, as explained in~\cite{regnier2018felix}. It reads
\begin{equation}
    g(\fvec{q},t=0) \propto \sum_k \exp\left[\frac{(E_k - \GCMExtraEneInit)^{2}}{2\GCMSigInit^2}\right] g_k^{\text{extra}}(\fvec{q})\ .
\label{eq:phi_init}
\end{equation}
The parameter $\GCMSigInit$ is a free parameter of the model that controls the energy spread of the wave packet. Given $\GCMSigInit$, we adjust $\GCMExtraEneInit$ to set the initial wave packet's energy relative to the non-extrapolated Potential Energy Surface (PES) defined by $V(\fvec{q})$. The wave-packet is absorbed at the borders of the deformation domain by adding an imaginary term in Eq.~\eqref{eq:theo:tdgcm:tdgcmgoa} to avoid spurious reflections~\cite{regnier2016fission}.

\subsubsection{Generator States}
\label{sec:theo:hfb}
We assume that all the many-body states $\ket{\idHFB{\fvec{q}}}$ of the PES are fermionic Bogoliubov vacuums. These product states of quasiparticles are defined as
\begin{equation}
    \ket{\idHFB{\fvec{q}}} \equiv \prod_{\mu} \aQP[\fvec{q}]{\mu}\ket{\idPV}\ ,
\end{equation}
where
\begin{equation}
    \aQP[\fvec{q}]{\mu} = \sum_k \big( U_{k\mu}^{(\fvec{q})*}\aPw{k} + V_{k\mu}^{(\fvec{q})*}\cPw{k} \big)\ .
\end{equation}
The operators $\aPw{k}$ and $\cPw{k}$ are the ladder operators associated with an arbitrary single-particle basis, and the complex numbers $U_{k\mu}^{(\fvec{q})}$ and $V_{k\mu}^{(\fvec{q})}$ define the quasi-particle state $\mu$ at point $\fvec{q}$. They are obtained through standard constrained HFB calculations with the constraints $\fvec{q}$ of Eq.~\eqref{eq:constraints} on the collective variables.

This work is a follow-up study of~\cite{schunck2014,schunck2015b,regnier2016fission}. Therefore, we make the exact same choices concerning the nuclear energy functional: the nuclear part of the total energy is computed from the Skyrme energy density functional (EDF) with the \texttt{Skm\textsuperscript{*}} \cite{bartel1982} parametrization; the pairing part of it is calculated from a surface-volume, density-dependent pairing force locally adjusted to reproduce the odd-even staggering of $^{240}$Pu, with a cut-off of 60 MeV; the Coulomb exchange part is computed at the Slater approximation. The HFB equation is solved by expanding the solutions in the stretched harmonic oscillator basis; see \cite{schunck2014} for numerical details. Practical calculations were performed with the \texttt{HFODD} solver~\cite{schunck2017solution}.

\subsubsection{Collective Inertia}
\label{sec:theo:inertia}

It is well-known that the GCM theory does not reproduce the asymptotic limit in the case of translational motion when considering only a collective subspace of time-even generator states $\fvec{q}$~\cite{baranger1978}.
In principle, this limitation can be remedied by doubling the number of degree-of-freedom via the introduction of conjugate momenta $\fvec{p}$ for each collective variable $\fvec{q}$~\cite{goeke1980}.
However, it has not been attempted in realistic calculations yet and remains to be verified~\cite{bender2020}.
Instead, it has become customary in fission calculations to use the adiabatic time-dependent Hartree-Fock-Bogoliubov (ATDHFB) approach to estimate the collective inertia, even though it breaks the internal consistency of the theory~\cite{verriere2020timedependent}.
Better methods are slowly becoming available~\cite{matsuo2000,hinohara2007}.
Since this work focuses on the impact of particle number projection techniques, we will simply use the GCM inertia at the cranking approximation (only the diagonal part of the RPA matrix is considered and time-odd terms are neglected) with derivatives computed locally (``perturbative'' cranking); see \cite{schunck2016microscopic} for additional details.

\subsection{Fission Fragments Distributions}
\label{sec:theo:proj}
In practical applications, fission models that rely on the calculation of a potential energy surface (e.g., the semi-classical random walk and Langevin approaches or the fully-microscopic TDGCM) cannot correctly describe the separation of the fissioning nucleus into two fragments that are simultaneously completely separated and do not interact anymore (other than through Coulomb), yet also excited. Instead, quantities such as fission fragment distributions are computed just before scission, which is defined, somewhat arbitrarily, based on several possible criteria; see discussions in \cite{schunck2016microscopic,younes2019}. 

\subsubsection{Population of Scission Configurations}
\label{sec:theo:proj:population}

Following common practice, we define scission configurations through the average value of the operator, $\Qneck$,  counting the number of nucleons in the neck. It is defined, as in Ref.~\cite{warda2002}, by
\begin{equation}
    \Qneck = e^{-(z-\zneck)^2/\aneck^2}\ .
\end{equation}
This expression contains two parameters. The neck location, $\zneck$, is the point between the two prefragments along the $z$-axis of the intrinsic reference frame where the local density is minimum. The position of the neck defines two prefragments: a point $(x,y,z)$ belongs to the ``left'' fragment if $z < \zneck$ and to the ``right'' fragment if $z \geq \zneck$. The dispersion, $\aneck$, is chosen to be equal to one nucleon. By convention, we consider that a configuration is located past scission, i.e., corresponds to two fully separated fragments, if
\begin{equation}
    \qneck \equiv \braket[\Qneck]{\idHFB{\fvec{q}}}{\idHFB{\fvec{q}}} \leq \qnecksciss\ ,
\end{equation}
where $\qnecksciss$ is a parameter. This allows us to define the scission configurations as the set of all the states $\ket{\idHFB{\fvec{q}}}$ such that (i) $\qneck > \qnecksciss$, and (ii) at least one of their neighbor is located past scission. 

In our two-dimensional calculations, the set of scission configurations can be parameterized by a single coordinate $\xi$. The population of scission configurations is then obtained by integrating the flux density $\phi(\xi, t)$, which reads in this specific case
\begin{equation}
    \phi(\xi, t) = J_0(\fvec{q}(\xi), t)\frac{\diff{q_1}}{\diff{\xi}} - J_1(\fvec{q}(\xi), t)\frac{\diff{q_0}}{\diff{\xi}}\ ,
\end{equation}
where $\fvec{J} = (J_0, J_1)$ is the current as defined in~\cite{regnier2016fission}.
We assume that the probability to exit through the point $\fvec{q}(\xi)$ is proportional to the time-integrated flux density, defined as
\begin{equation}\label{eq:raw_flux}
    F(\xi) = \lim_{t\rightarrow\infty}\int_{\tau=0}^{\tau=t}\diff{\tau} \phi(\xi, \tau)\ .
\end{equation}

\subsubsection{Expression of the yields}
\label{sec:theo:proj:yields}

Previous studies using the TDGCM framework to predict fission fragment yields accounted for the width of the particle-number probability distribution at a single scission configuration, as well as a possible experimental mass resolution (to compare with experimental pre-neutron yields), by convoluting the time-integrated flux density of Eq.~\eqref{eq:raw_flux} with a Gaussian function~\cite{regnier2016fission,tao2017,regnier2019from,zhao2019,zhao2019a}.
The mean of such a Gaussian is typically set to the average number of particles in the right fragment at the scission configuration $\ket{\phi(\fvec{q}(\xi))}$, while its standard deviation is a free parameter.
The major novelty of this work is the determination of the fission fragment mass and charge yields, $Y(\Zfrag, \Afrag)$ by computing the exact probabilities for each fragments $\Zfrag, \Afrag$ using a particle-number projection-based technique rather than a convolution with a Gaussian.

We seek to calculate the probability $\cprob{\ZR,\NR}{Z,N}$ to measure $\ZR$ charges and $\NR$ neutrons in the right fragment at scission given a compound system with $Z$ protons and $N$ neutrons. In our approach, it is given by
\begin{equation}
\label{eq:pzrnr}
    \cprob{\ZR,\NR}{Z,N} \propto
      \int \diff{\xi}\ F(\xi)\ \cprob{\ZR,\NR}{Z, N, \fvec{q}(\xi)}\ ,
\end{equation}
where $\cprob{\ZR,\NR}{Z, N, \fvec{q}(\xi)}$ is the probability, whose determination is presented in \ssecref{sec:theo:proj:PNP}, associated with a right fragment having $\ZR$ protons and $\NR$ neutrons, only considering the part of the compound system's state at $\fvec{q}(\xi)$ having $Z$ protons and $N$ neutrons. Once this quantity is inserted in Eq.~\eqref{eq:pzrnr} and after a proper normalization, we recover the fission fragments yields, in percent, assuming that the system splits in two fragments
\begin{multline}
    Y(\Zfrag,\Nfrag) = 100 \times \big[\cprob{\ZR=\Zfrag,\NR=\Nfrag}{Z,N} \\
                      +\cprob{\ZR=Z-\Zfrag,\NR=N-\Nfrag}{Z,N} \big]\ .
    \label{eq:yields}
\end{multline}

We could use the same procedure to determine one-dimensional yields. For example, the charge distributions would read
\begin{multline}
    Y(\Zfrag) = 100 \times \big[ \cprob{\ZR=\Zfrag}{Z,N} \\
                      +\cprob{\ZR=Z-\Zfrag}{Z,N} \big]\ ,
    \label{eq:chargeyields}
\end{multline}
where the probability $\cprob{\ZR}{Z,N}$, associated with $\ZR$ protons in the right fragment at scission given $Z$ and $N$, is
\begin{equation}
\label{eq:pzr}
    \cprob{\ZR}{Z,N} 
    \propto  
    \int \diff{\xi}\ F(\xi)\ \cprob{\ZR}{Z,N,\fvec{q}(\xi)}\ ,
\end{equation}
and the marginalized probability $\cprob{\ZR}{Z,N,\fvec{q}}$ is given by
\begin{equation}
    \cprob{\ZR}{Z,N,\fvec{q}}  = \sum_{n=0}^{N}\cprob{\ZR, \NR=n}{Z,N,\fvec{q}}\ .
    \label{eq:chargeproba}
\end{equation}
However, all the distributions of interest can alternatively be obtained directly from $Y(\Zfrag,\Nfrag)$ through the relations
\begin{align}
    Y(\Zfrag, \Afrag) &= Y(\Zfrag, \Nfrag = \Afrag-\Zfrag) \\
    Y(\Zfrag) &= \sum_{n=0}^{N} Y(\Zfrag, \Nfrag = n) \\
    Y(\Afrag) &= \sum_{n=0}^{} Y(\Zfrag=\Afrag-n, \Nfrag=n)\ .
\end{align}

Therefore, the essential quantity to compute the fission yields is the probability distribution $\cprob{\ZR,\NR}{Z, N, \fvec{q}}$.

\subsubsection{Particle Number Projection}
\label{sec:theo:proj:PNP}

In this work, we estimate the distribution of probability $\cprob{\ZR,\NR}{Z,N,\fvec{q}}$ based on particle-number projection techniques. 
Particle-number projection (PNP) was originally introduced to restore the number of particles in superfluid systems that spontaneously break the particle number symmetry~\cite{mang1975,blaizot1985,bender2003,ring2004,schunck2019}.
Following Refs~\cite{simenel2010particle,scamps2015superfluid,verriere2019number}, we use PNP to estimate the probability $\cprob{\ZR,\NR}{Z,N,\fvec{q}}$.
For a fissioning nucleus described by the many-body state $\ket{\phi(\fvec{q})}$, this value can be interpreted as the probability to measure $\Zfrag$ protons and $\Nfrag$ neutrons in the right fragment in the component of $\ket{\phi(\fvec{q})}$ with $Z$ protons and $N$ neutrons.

We do not consider isospin mixing in this work, i.e.,
\begin{equation}
    \ket{\idHFB{\fvec{q}}} = \ket{\idHFB{\fvec{q}}}_{\rm neut.} \otimes \ket{\idHFB{\fvec{q}}}_{\rm prot.}\ .
\end{equation}
Thus, the probabilities $\cprob{\ZR,\NR}{Z,N,\fvec{q}}$ can be decomposed accordingly,
\begin{equation}
    \cprob{\ZR,\NR}{Z,N,\fvec{q}}
    = 
    \cprob{\NR}{N,\fvec{q}} \times \cprob{\ZR}{Z,\fvec{q}}\ .
\label{eq:double_proj}
\end{equation}
The probability distributions $\cprob{\ZR}{Z,\fvec{q}}$ and $\cprob{\NR}{N,\fvec{q}}$ both derive from a double projection of the scission configuration $\ket{\phi(\fvec{q})}$
\begin{equation}\label{eq:doubleproj}
    \cprob{\ZR}{Z,\fvec{q}} =
      \frac{ \braket[\ProjZR(\ZR) \ProjZ(Z)]{\idHFB{\fvec{q}}}{\idHFB{\fvec{q}}} }{\braket[\ProjZ(Z)]{\idHFB{\fvec{q}}}{\idHFB{\fvec{q}}}}\ .
\end{equation}
The operator $\ProjZ(Z)$ projects the state onto the states' eigenspace with a good total proton number $Z$.
We implemented this projector in its standard gauge angle integral form
\begin{equation}
    \ProjZ(Z) = 
    \frac{1}{2\pi}\int_{0}^{2\pi} \diff{\theta}\ e^{i\theta(\op{Z}-Z)}\ .
\end{equation}
Similarly, the operator $\ProjZR(\ZR)$ projects onto the eigenspace of states with a good number of protons $\ZR$ in the right half-space.
We recall that the right half-space corresponds to the set of spatial coordinates whose component along the $z$-axis is greater than $\zneck$, the position of the neck.
Its expression reads
\begin{equation}
    \ProjZR(\ZR) = 
    \frac{1}{2\pi}\int_{0}^{2\pi} \diff{\theta}\ e^{i\theta(\opZR-\ZR)},
\end{equation}
where $\opZR$ counts the number of protons in the right half-space.
This operator can be expressed from the proton creation ($\cPr[\rm p]{\fvec{r}, \sigma}$) and annihilation ($\aPr[\rm p]{\fvec{r}, \sigma}$) operators as
\begin{align}
    \op{Z}_{\rm R} &= 
    \sum_{\sigma=\downarrow,\uparrow} 
    \int_{x}\int_{y}
    \int_{z=\zneck}^{+\infty} \diff{\fvec{r}}\
    \cPr[\rm p]{\fvec{r}, \sigma}
    \aPr[\rm p]{\fvec{r}, \sigma}\ .
    \label{eq:counting_op}
\end{align}
We define the projectors on the neutrons number, in the full space ($\ProjN(N)$) as well as in the right half-space ($\ProjNR(\NR)$) in a similar way,
and we follow the same procedure to calculate the neutron probability
\begin{equation}\label{eq:doubleprojN}
    \cprob{\NR}{N,\fvec{q}} =
      \frac{ \braket[\ProjNR(\NR) \ProjN(N)]{\idHFB{\fvec{q}}}{\idHFB{\fvec{q}}} }{\braket[\ProjN(N)]{\idHFB{\fvec{q}}}{\idHFB{\fvec{q}}}}\ .
\end{equation}

In our calculations, we determine the numerator of Eqs.~\eqref{eq:doubleproj} and~\eqref{eq:doubleprojN} using the gauge angle integrals based on a Fomenko quadrature~\cite{fomenko1970} using 41 integration points.
Instead of explicitely calculating the corresponding denominator, we directly normalize the distributions using
\begin{align}
  \braket[\ProjZ(Z)]{\idHFB{\fvec{q}}}{\idHFB{\fvec{q}}} &=
    \sum_z \braket[\ProjZR(z) \ProjZ(Z)]{\idHFB{\fvec{q}}}{\idHFB{\fvec{q}}} \\
  \braket[\ProjN(N)]{\idHFB{\fvec{q}}}{\idHFB{\fvec{q}}} &=
    \sum_n \braket[\ProjNR(n) \ProjN(N)]{\idHFB{\fvec{q}}}{\idHFB{\fvec{q}}}\ .
\end{align}

\section{Application}
\label{sec:res}

This section summarizes our results for the thermal neutron-induced fission of $^{235}$U and $^{239}$Pu. In \ssecref{sec:res:PES}, we discuss the static fission properties related to the potential energy surfaces and the properties of prefragments at scission. In \ssecref{sec:res:projection}, we show the primary fission fragments mass and charge yields obtained by combining the TDGCM+GOA collective dynamics with PNP. We then compare the projected yields with results obtained from assuming several analytical distributions in \ssecref{sec:res:analytic}, and assuming different criteria for the scission line in \ssecref{sec:res:qn}. Finally, the evolution of both the two-dimensional yields and their one-dimensional reductions as a function of the energy of the incident neutron is illustrated in \ssecref{sec:res:En}.

\subsection{Static properties}
\label{sec:res:PES}

\begin{figure}[!ht] 
\begin{center} 
\includegraphics[width=\columnwidth]{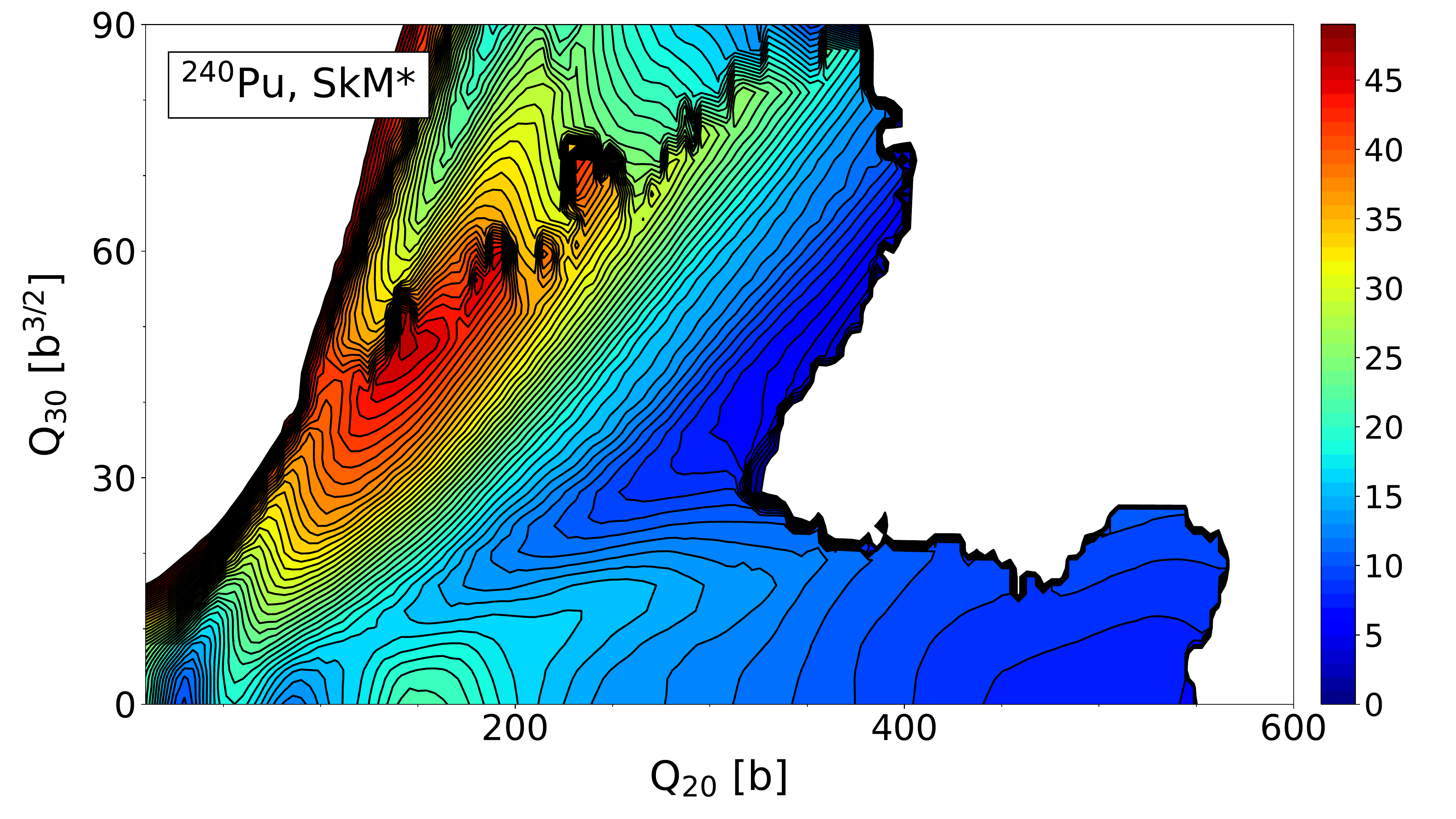}
\includegraphics[width=\columnwidth]{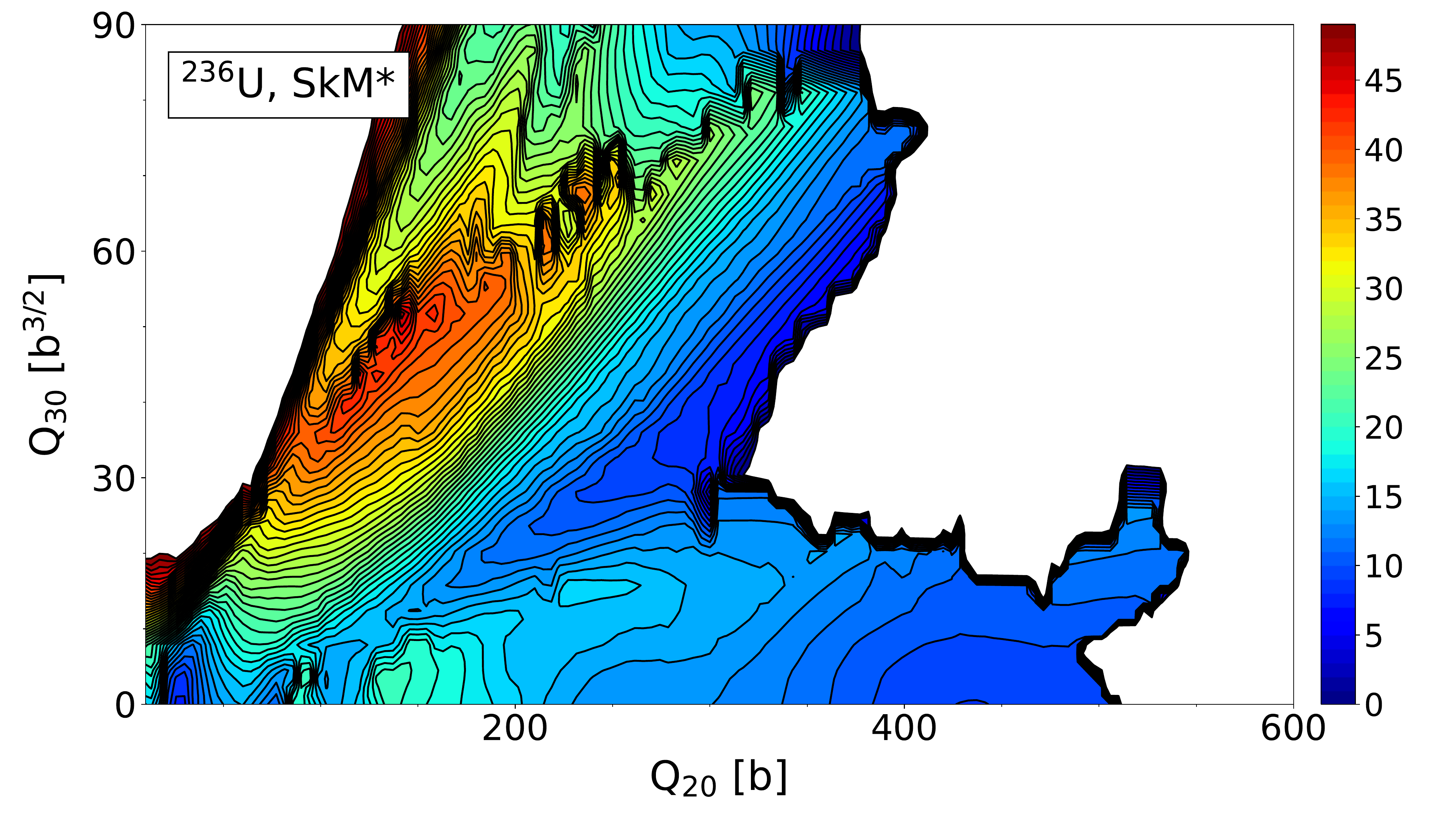}
\caption{\label{fig:PES} Potential energy surface for $^{240}$Pu (top) and $^{236}$U (bottom) as a function of the axial quadrupole and axial octupole moments.} 
\end{center} 
\end{figure}

While the potential energy surfaces used in this work have already been presented in other publications -- \cite{schunck2014} for $^{240}$Pu and \cite{schunck2020b} for $^{236}$U -- we recall them for the sake of completeness. Figure \ref{fig:PES} thus shows the total potential energy  $V(\fvec{q})$ entering the collective Schr\"odinger equation of Eq.~\eqref{eq:theo:tdgcm:tdgcmgoa} for both $^{240}$Pu (top) and $^{236}$U (bottom) as a function of the axial quadrupole and octupole moments. The PES is characterized by a ground-state at around $Q_{20} \approx 30$~b, a fission isomer at $Q_{20} \approx 85$~b, and the main large fission valley opening up beyond the second barrier. Note that triaxiality is included in this calculation, but plays a role only near the first barrier; see \cite{schunck2014} for a discussion. Additional details about the resolution of the HFB equation, such as the characteristics of the harmonic oscillator basis or the convention for the multipole operators used as constraints can be found in that same reference.

An HFB generator state $\ket{\idHFB{\fvec{q}}}$ is associated with each point of the PES of \figref{fig:PES}. The scission configurations are defined based on the procedure outlined in Sec.~\ref{sec:theo:proj}. For each state in the scission region, we identify the neck position $\zneck$, which is used to compute the average charge, number of neutron, and mass, according to
\begin{align}
\label{eq:meanvalues}
    \qavg{\op{Z}_{\rm R}} =
      \int_{-\infty}^{+\infty} \diff{x}
      \int_{-\infty}^{+\infty} \diff{y}
      \int_{\zneck}^{+\infty} \diff{z}
        \ \rho_{\rm p}(\fvec{r}) , \\
    \qavg{\op{N}_{\rm R}} =
      \int_{-\infty}^{+\infty} \diff{x}
      \int_{-\infty}^{+\infty} \diff{y}
      \int_{\zneck}^{+\infty} \diff{z}
      \ \rho_{\rm n}(\fvec{r}) , \\
    \qavg{\op{A}_{\rm R}} =
      \int_{-\infty}^{+\infty} \diff{x}
      \int_{-\infty}^{+\infty} \diff{y}
      \int_{\zneck}^{+\infty} \diff{z}
      \ \rho_{\rm 0}(\fvec{r}) ,
\end{align}
where $\rho_{\rm n}$ and $\rho_{\rm p}$ are respectively the neutron and proton density distributions and $\rho_0 = \rho_{\rm n}+\rho_{\rm p}$ is the isoscalar (total) density. Since the HFB wave function is not an eigenstate of the operator of Eq.~\eqref{eq:counting_op} counting the number of particles in the right fragment, the fluctuations $\langle \opZR^{2} \rangle$ and $\langle \opAR^{2} \rangle$ are non-zero and could be in principle computed as well.

Projection techniques give us a much more complete view of the content in particle number in the fragments.
For each scission configuration of the two PES we performed the projection on the total number of protons and neutrons as well as on the number of protons and neutrons in the right half-space following Eq.~\eqref{eq:doubleproj}.
The projection on the fragment particle numbers was only performed for eigenvalues $\ZR$ and $\NR$ close to the mean values defined by Eq.~\eqref{eq:meanvalues}, where the probability is not negligible. In practice we considered in each scission configuration the set $\lfloor \langle \opZR \rangle \rfloor - 20 \leq \ZR < \lfloor \langle \opZR\rangle \rfloor + 20$ and $\lfloor \langle \opNR\rangle \rfloor - 20 \leq \NR < \lfloor \langle \opNR \rangle \rfloor + 20 $, hence a total of 41 projected values for protons and for neutrons. 
The notation $\lfloor X \rfloor$ indicates the integer part of the real number $X$.

\begin{figure}[!ht] 
\begin{center} 
\includegraphics[width=\columnwidth]{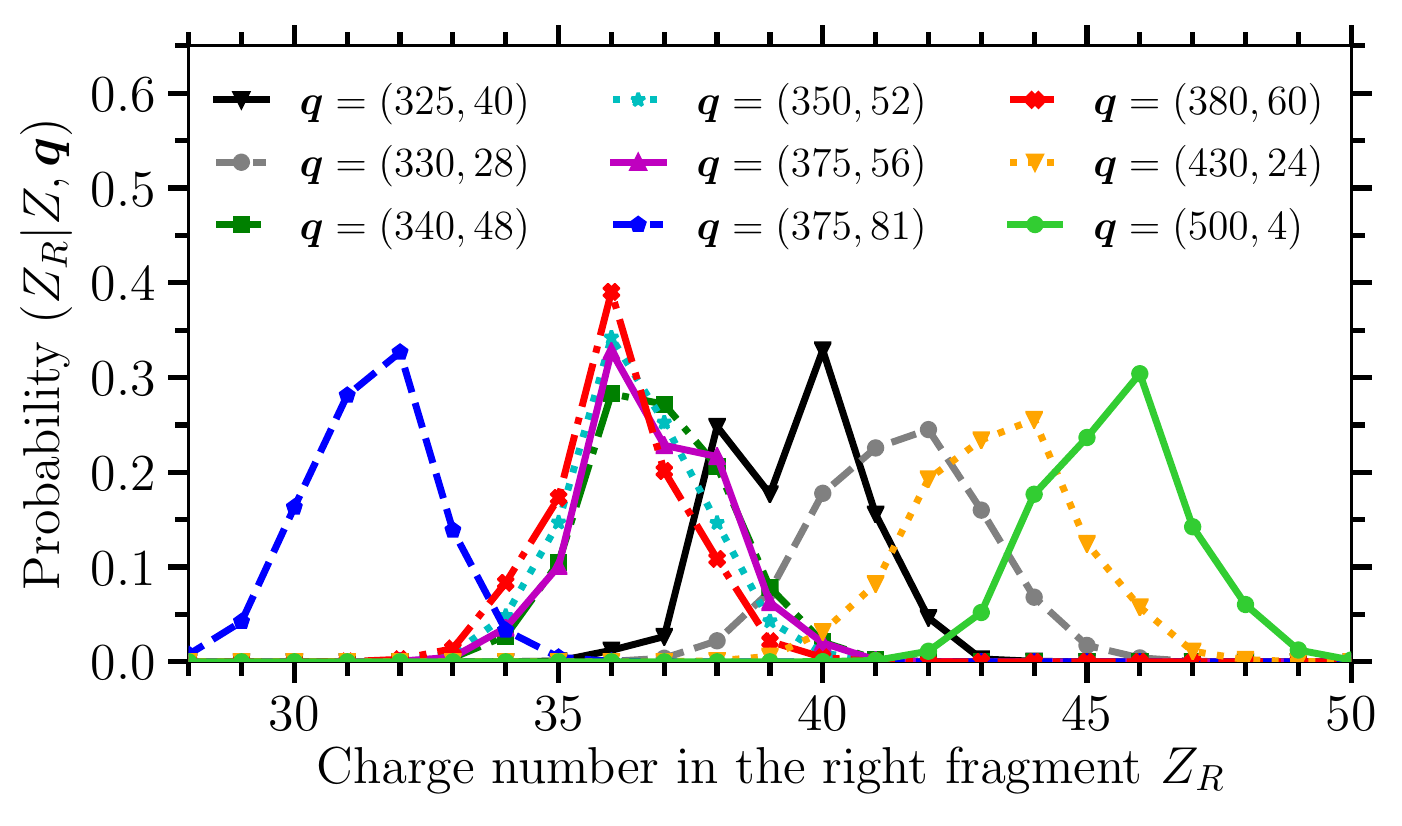}
\caption{\label{fig:proj} Probability $\cprob{\ZR}{Z,\fvec{q}}$ of having $\ZR$ protons in the right fragment given the total number of protons $Z$ as a function of $\ZR$. We show such a probability distribution for a set of scission configurations $\fvec{q}$ in $^{236}$U. The deformations $\fvec{q}$ are in b ($q_{20}$) and b$^{3/2}$ ($q_{30}$), where b = barn.} 
\end{center} 
\end{figure}

Figure \ref{fig:proj} presents an example of these probabilities distributions for the case of a few scission configurations in $^{236}$U, the locations $\fvec{q}$ of which are indicated in the legend. More specifically, it shows the probability $\cprob{\ZR}{Z,\fvec{q}}$ as a function of the charge number $\ZR$ for the fixed value $Z=92$ of the total number of proton in the fissioning nucleus. These results highlight two important features brought about by projection: (i) the curves are not necessarily symmetric around the mean value and (ii) there can be odd-even staggering effects, as in the case of the configuration $(q_{20},q_{30}) = (325\;\mathrm{b}, 40\;\mathrm{b}^{3/2})$. These two features are absent by construction when considering empirical dispersion laws such as Gaussian folding.

\subsection{Fragment distributions}
\label{sec:res:projection}

In this section, we show the primary mass and charge distributions for the two important cases of the low-energy, neutron-induced fission of $^{235}$U and $^{239}$Pu. 
Throughout this section, FELIX calculations were performed with the GCM collective inertia, the metric $\gamma(\fvec{q})$ and the GCM zero point energy correction, and the frontier was defined by $\qnecksciss = 4.0$. In Eq.~\eqref{eq:phi_init}, the initial collective wave packet had a width $\GCMSigInit = 0.5$ MeV. The collective dynamics with FELIX was characterized by a time step of $\Delta t = 2\times 10^{-4}$ zs and we always simulated the dynamics up to a time $t=20$ zs.

For the two actinides under consideration, there are considerably fewer measurements on charge distributions $Y(Z)$ than on mass distributions $Y(A)$. 
Yet charge distributions are important tests for theory since they can exhibit an odd-even effect, namely, the yield of even-$Z$ elements is higher than that of odd-$Z$ \cite{ehrenberg1972,amiel1975,mariolopoulos1981,bocquet1989,gonnenwein1992}. 
Note that an experimental determination of such an effect requires a sufficiently good resolution in the detection of the fragment charge. 
Typically, for the case of neutron-induced fission of $^{235}$U at thermal energy, odd-even effects were not reported in \cite{reisdorf1971}. According to the authors, the experimental technique used as well as the hypothesis assumed in the data analysis could smooth out the odd-even structures in this work.
On the contrary, the experiment \cite{lang1980nuclear} claims a resolving power of $Z/\Delta Z=45$ (full-width 1/10 maximum) for the charge $Z=40$ at the maximum of the light peak.
With such resolution, the resulting yields present a strong odd-even staggering effect.
Such effect was also observed for various fissioning systems in the high precision measurements based on inverse kinematics reactions at higher excitation energies~\cite{schmidt2000,ramosIsotopic2018,chatillonExperimental2019}.

\begin{figure}[!ht] 
\begin{center} 
\includegraphics[width=\columnwidth]{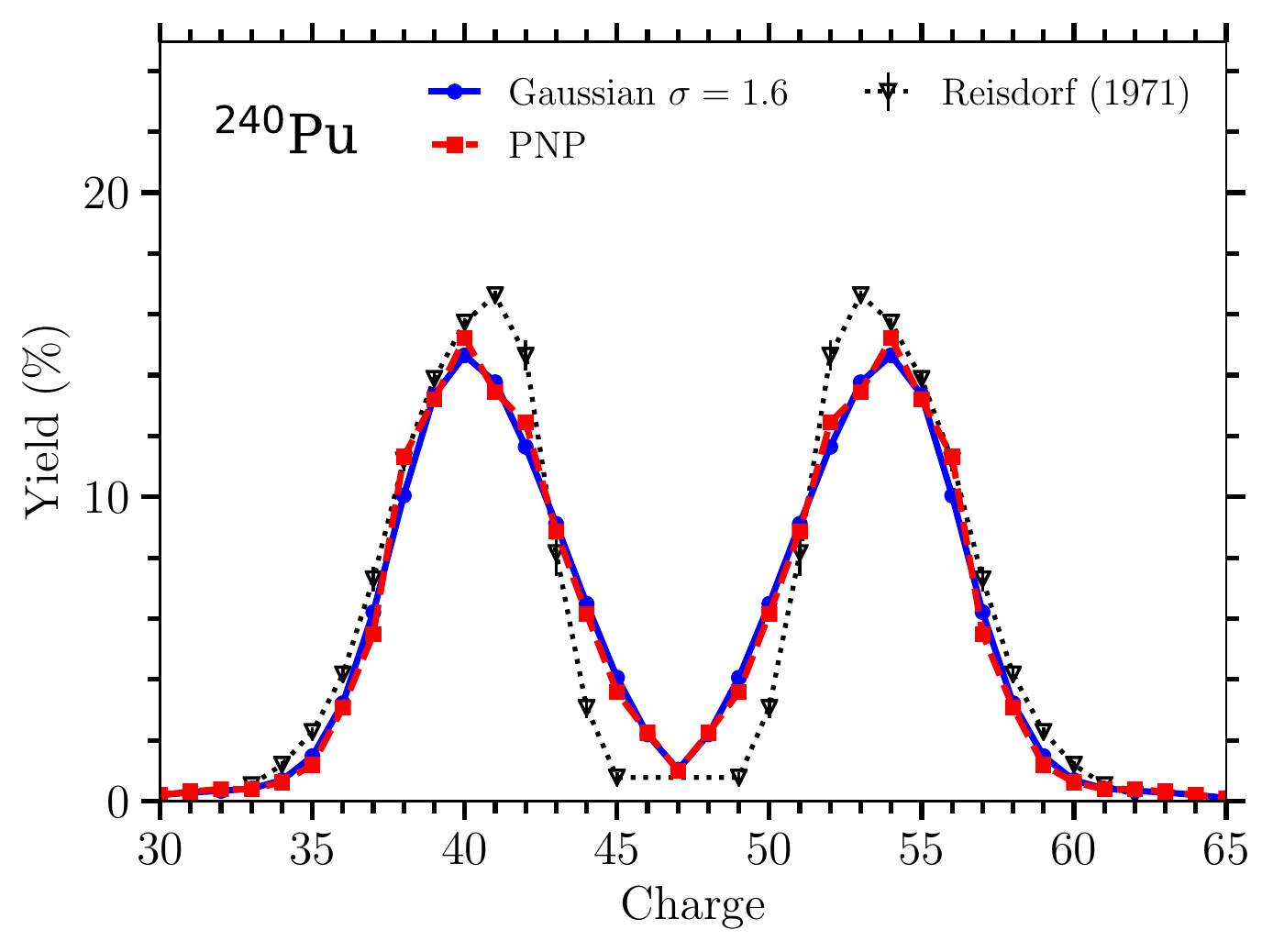}
\includegraphics[width=\columnwidth]{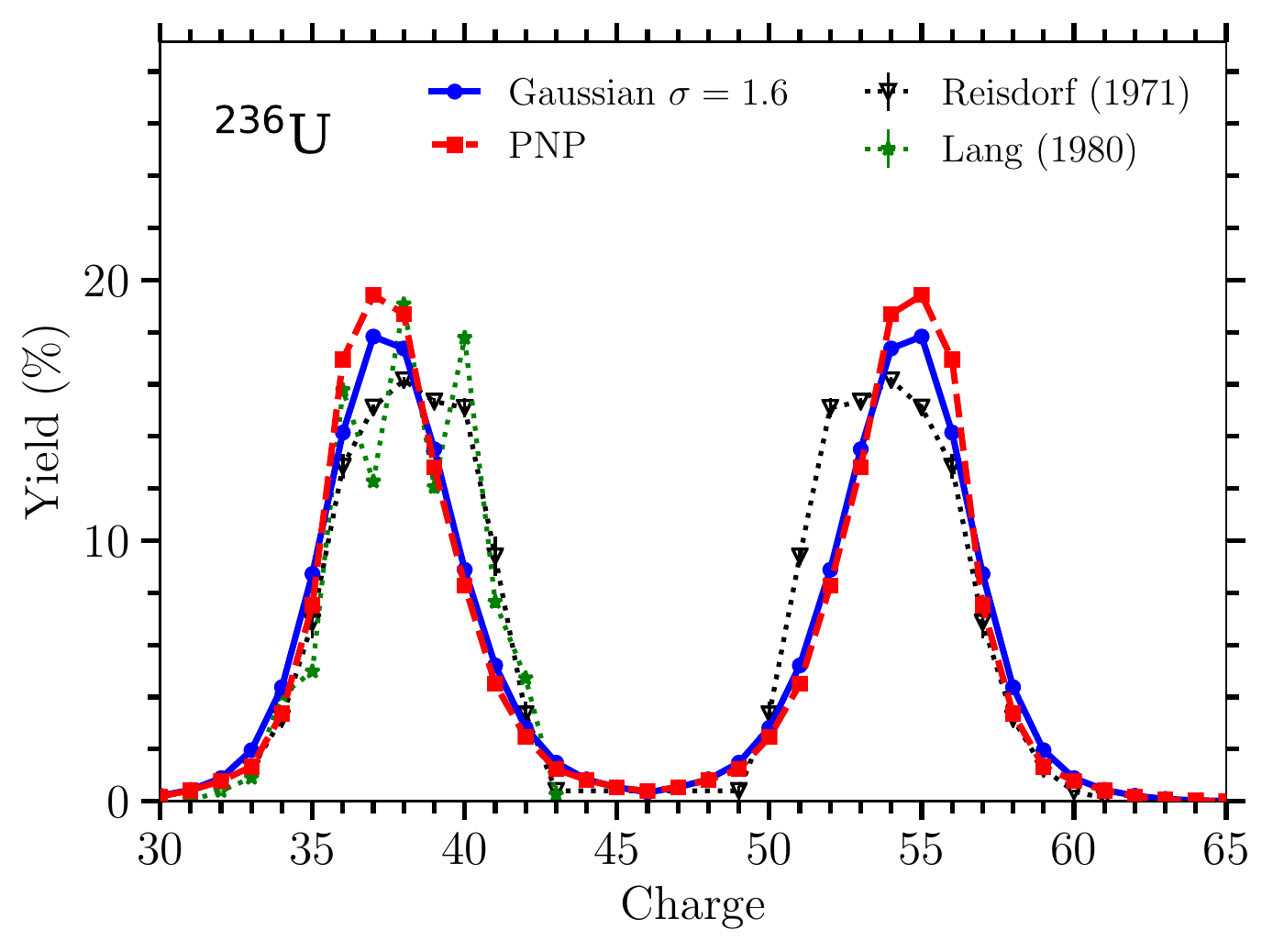}
\caption{\label{fig:YZ_exp} Charge distribution $Y(Z)$ in $^{240}$Pu (top) and $^{236}$U (bottom) as a function of $Z$. The yields obtained with PNP are compared to a Gaussian convolution of the raw flux with $\sigma=1.6$ and to the experimental data from \cite{reisdorf1971,lang1980nuclear}; see text for additional details.}
\end{center} 
\end{figure}

In \figref{fig:YZ_exp}, we show the charge distributions for the neutron-induced fission $^{239}$Pu(n,f) and $^{235}$U(n,f) reactions, at an excitation energy of 1 MeV above the first fission barrier. 
For these fissile isotopes it should correspond to an incident neutron energy of $E_n \simeq 1$ MeV. 
For both systems, the overall agreement with data is satisfactory, especially considering the lack of precise experimental measurements of charge distributions for the $^{240}$Pu. In the case of $^{236}$U, it is worth noting that our calculations do not predict odd-even effects in the charge yields. We will discuss this in more details in Sec. \ref{sec:res:qn}.

\begin{figure}[!ht] 
\begin{center} 
\includegraphics[width=\columnwidth]{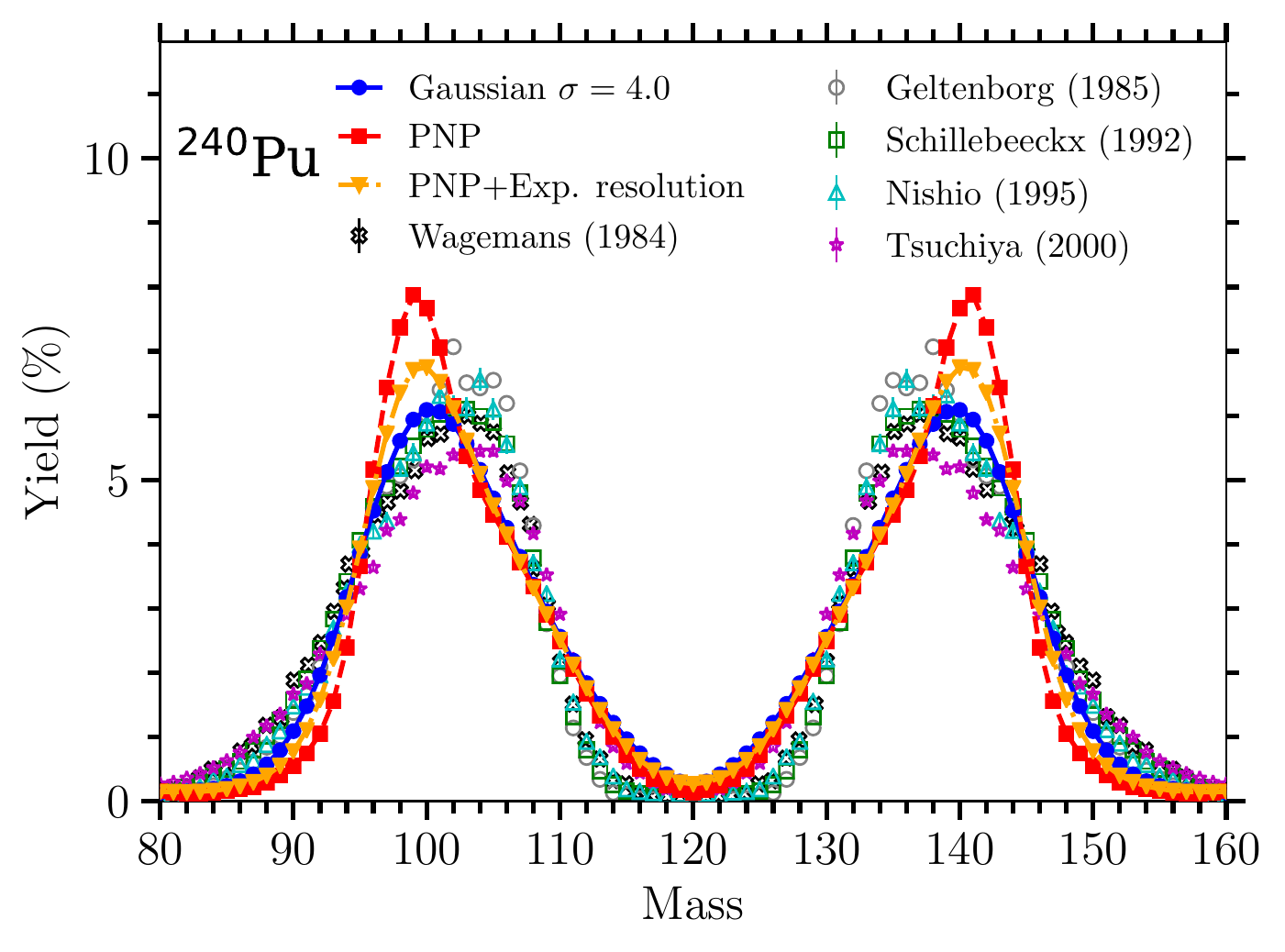}
\includegraphics[width=\columnwidth]{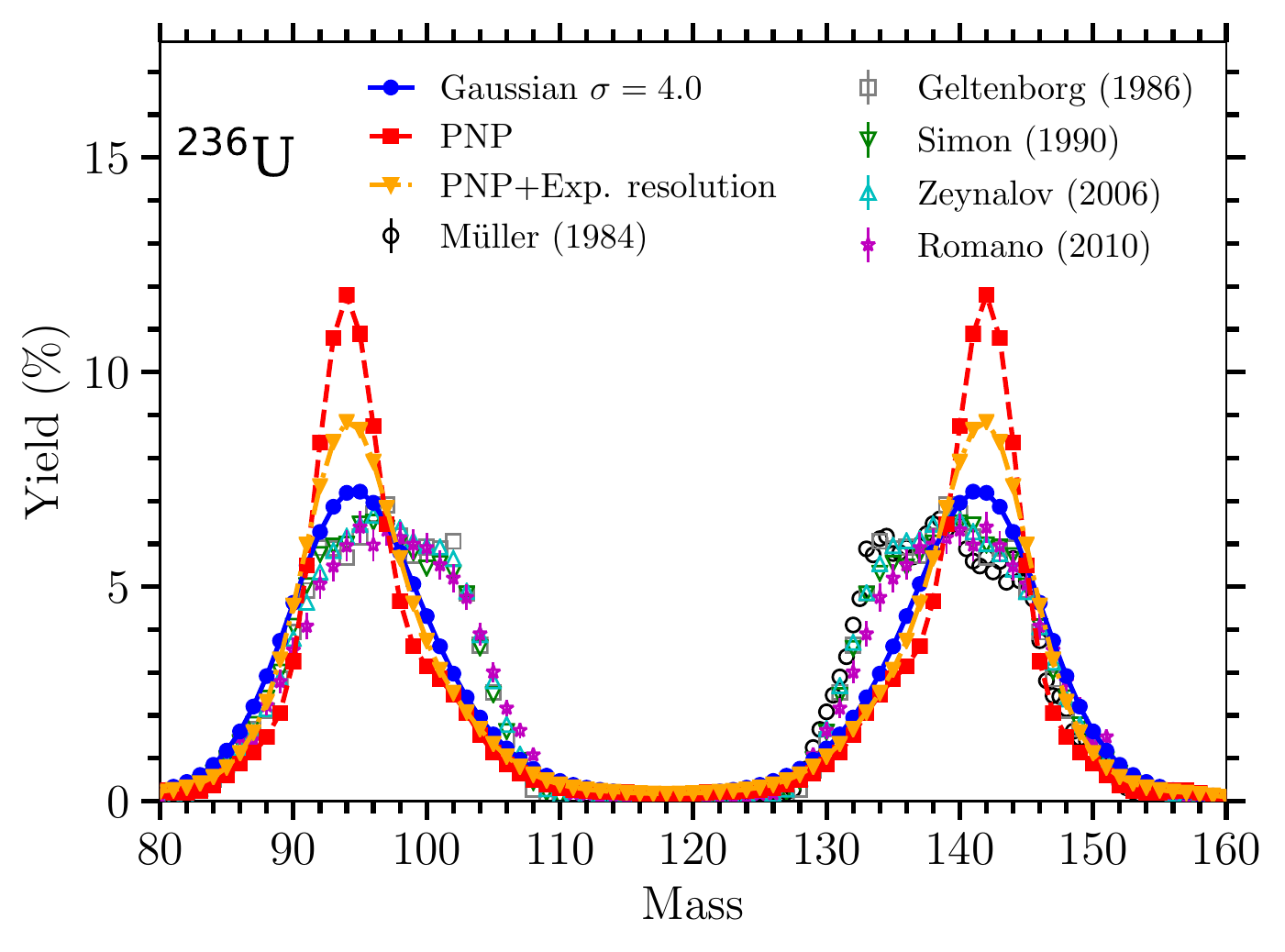}
\caption{\label{fig:YA_exp} Mass distribution $Y(A)$ in $^{240}$Pu (top) and $^{236}$U (bottom) as a function of $A$. Results obtained after PNP are compared to a Gaussian convolution of the raw flux with $\sigma=4.0$ and to experimental data from~\cite{wagemansComparison1984,geltenbortPrecision1986,schillebeeckxComparative1992,nishioMeasurement1995a,tsuchiyaSimultaneous2000,mullerFragment1984,simonPulse1990,zeynalovInvestigation2006,romanoFission2010}. The curve labeled PNP+Exp. resolution gives the PNP results with an additional convolution by a Gaussian with a FWHM of 5.5 mass units to account for the typical mass resolution of experiments; see text for additional details.
%
} 
\end{center} 
\end{figure}

In \figref{fig:YA_exp}, we show the mass distributions for these same isotopes. The agreement with data is not as good, especially in the case of $^{235}$U(n,f). However, we must bear in mind that these experimental primary mass yields do not have a perfect mass resolution. The mass of the (primary) fragments is typically deduced from the measurement of the kinetic energy of the fission fragments in ionisation chambers. This technique implies a mass resolution of roughly FWHM=5.5 mass units for A~\cite{gook2018prompt}. Therefore, we also show in \figref{fig:YA_exp} the results of the calculation when accounting for such a mass resolution. 

In both Figs.~\ref{fig:YZ_exp} and \ref{fig:YA_exp}, we compare the theoretical calculations with PNP with the standard method of Gaussian folding. Note that the width $\sigma$ of the folding must be different for protons and for masses as the overall proton density is smaller than the total one. As a result, most of the (theoretical) uncertainty on the proton number can be attributed to pairing effects in each prefragment. As seen in \figref{fig:proj}, the average dispersion around the mean value is of the order of 1.6 indeed. In contrast, neutrons are much less localized in the prefragments and contribute rather significantly to the overall neck between the prefragments \cite{younes2011,schunck2014,sadhukhan2017}. The number of particles in the neck used to define scission configuration in our calculations, $\qnecksciss = 4.0$, is larger than the typical dispersion in particle number one could expect from pairing effects and thus dominates. This justifies fixing $\sigma = 4$ for the Gaussian folding of mass yields. It is worth mentioning that for this reference calculation at $\qnecksciss = 4.0$, Gaussian folding is a very good approximation to the exact PNP result. In Sec.~\ref{sec:res:analytic}, we will evaluate the performance of other analytical probability laws, and in Sec.~\ref{sec:res:qn}, we will further discuss the dependency of the results on the definition of the frontier.

\begin{figure}[!ht] 
\begin{center}
\includegraphics[width=\columnwidth]{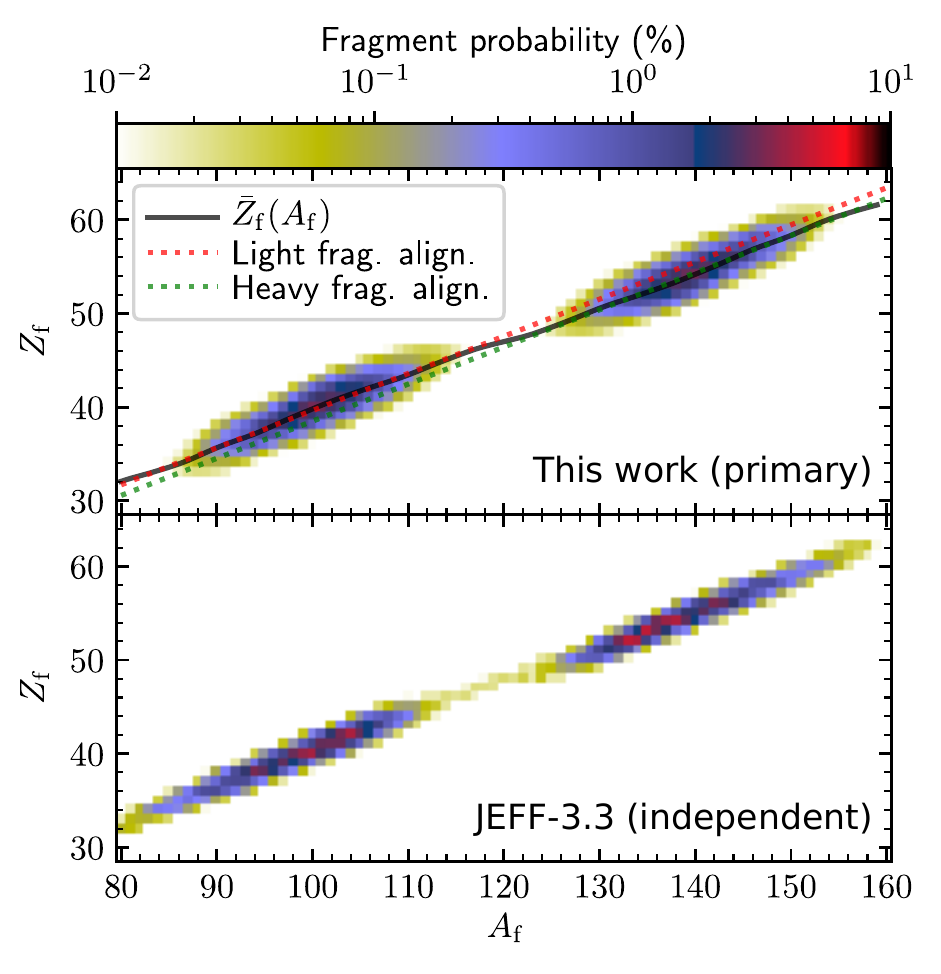}
\caption{\label{fig:YZA_pdf} Two-dimensional isotopic fission yields $Y(Z, A)$ for the reaction $^{239}$Pu(n,f) for an incoming neutron energy of $E_n=1.0$ MeV using PNP in the fragments. Our results predict a deviation from the Unchanged Charge Distribution (UCD) approximation, called the charge-polarization of the fission fragment distributions, in the primary yields. To give a point of reference, we also show the independent two-dimensional yields evaluated in the JEFF-3.3 library. Note that these yields account for the mass number of the fragments after emission of the prompt neutrons (contrary to our predictions that give the primary fragment yields).
}
\end{center}
\end{figure}

Beyond enhancing the robustness of our approach by eliminating the (somewhat) arbitrary width of the Gaussian folding, PNP also allows determining more realistic two-dimensional isotopic distribution $Y(Z,A)$. Such distributions are especially important when simulating the deexcitation of fission fragments \cite{becker2013}. The results for the particular case of $^{240}$Pu are presented in~\figref{fig:YZA_pdf}. 
To our knowledge, no experimental data can be directly compared to these pre-neutron emission yields. 
To give some reference point, we therefore the represent the two-dimensional independent yields (after the emission of the prompt neutrons) evaluated in the JEFF-3.3 evaluated data library~\cite{plompenJoint2020}.
We show that the PNP approach predicts a significant width for the light and heavy peaks of the two-dimensional fission yields. This width is larger than the data of JEFF-3.3. Currently, it is not clear whether this reduction of the width could be totally explained by the prompt neutron emission.

From $Y(\Zfrag,\Afrag)$ we can extract the charge polarization $\Delta \Zfrag$ which measures the deviation of the most probable charge in a fragment of given mass to the Unchanged Charge Distribution (UCD) approximation~\cite{wahl2002systematics},
\begin{equation}
    \Delta \Zfrag(\Afrag) = \bar{\Zfrag}(\Afrag) - \Afrag \frac{Z}{A},
\end{equation}
with
\begin{equation}
    \bar{\Zfrag}(\Afrag) = \frac{\sum_{\Zfrag} \Zfrag \times Y(\Zfrag, \Afrag)}{\sum_{\Zfrag} Y(\Zfrag, \Afrag)}.
\end{equation}
For the thermal neutron induced-fission induced on $^{239}$Pu, a large corpus of experimental data agree on a charge polarization of the order of -0.5 charge unit for the heavy fragments~\cite{schmittFission1984,bailIsotopic2011a}. We find in this work a consistent value of -0.58 on average on the fragments in the heavy peak ($\Afrag\in[130,150]$).
A qualitative explanation of this polarization effect, already proposed in 1966~\cite{norenbergTheory1966}, is the asymmetry energy term in the liquid drop formula of the deformation energy. However, no such pre-neutron experimental data were available at the time. As illustrated in Ref.~\cite{caamanoCharacterization2015}, classical models are able to explain the smooth trend of the charge polarization favoring less protons in the heavy fragment.
On top of that, shell and pairing effects brings additional structures to the charge polarization measured in the fission of actinides in general~\cite{naikSystematics1997,naikCharge2007}.

\begin{figure}[!ht]
    \centering
    \includegraphics[width=\columnwidth]{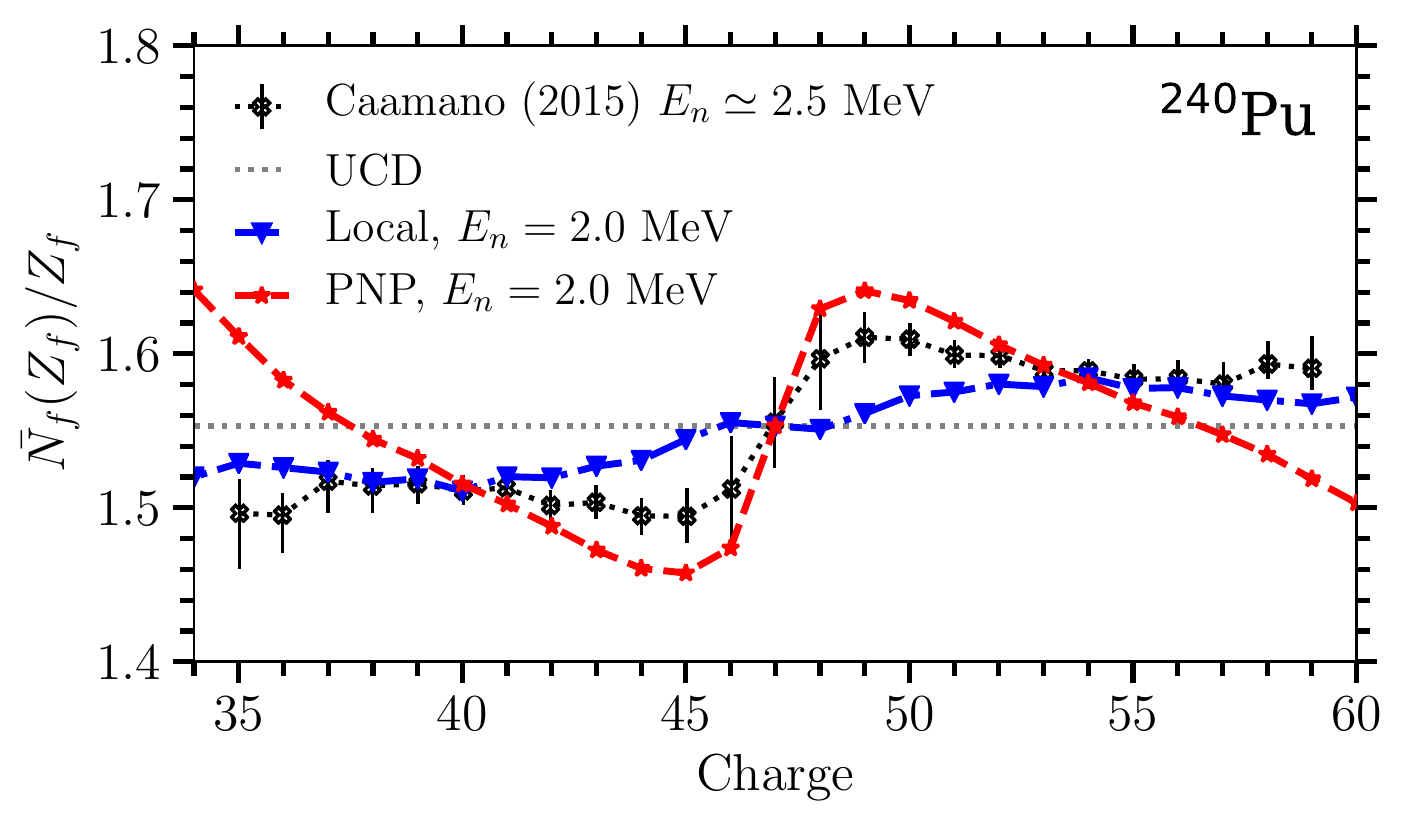}
    \caption{Average neutron excess of fragments produced in the fission of $^{240}$Pu. PNP results at  $E_n=2.0$ MeV are compared with the local discrete approach of \ref{sec:res:analytic} and with the experimental data of Ref.~\cite{caamanoCharacterization2015}.}
    \label{fig:pu240_neutron_excess}
\end{figure}

An equivalent probe for this phenomenon is the neutron excess defined as
\begin{equation}
\frac{\bar{\Nfrag}(\Zfrag) }{\Zfrag} 
= 
\frac{\sum_{\Nfrag} \Nfrag/\Zfrag \times Y(\Zfrag, \Nfrag)}{\sum_{\Nfrag} Y(\Zfrag, \Nfrag)}.
\label{eq:n_excess}
\end{equation}
We compare in Fig.~\ref{fig:pu240_neutron_excess} the predictions of the TDGCM dynamics with PNP to the experimental data obtained with the VAMOS spectrometer~\cite{caamanoCharacterization2015}. The experimental data comes from an inverse kinematic experiment with an equivalent incident neutron energy of $\simeq$ 2.5 MeV and a standard deviation of 6 MeV on this energy. To obtain an approximate description of this entrance channel, we performed our calculation with an initial energy of 2 MeV above the first fission barrier. 
The shape of the neutron excess obtained, especially the bump when going from the light to the heavy fragment and the overall order of magnitude in the light and heavy fragments, reproduces well this experimental data set.
At high asymmetries we notice significant deviations from experiment. 
To see the impact of projection on the quality of our prediction we also show the results obtained when the two-dimensional mass and charge yields are simply estimated with the local discrete approach of \ref{sec:res:analytic}. Particle number projection turns out to be an important ingredient for reproducing the neutron excess.

\subsection{Comparison with various analytical distributions}
\label{sec:res:analytic}

Many studies of fission fragment distributions in the framework of the TDGCM+GOA approximate the uncertainty on particle number at scission, or equivalently, particle transfer effects, by simple Gaussian folding of the mean value of particles for each scission configuration \cite{younes2012a,regnier2016fission,tao2017,regnier2019,zhao2019,zhao2019a}. 
This corresponds to setting the quantity 
$
\cprob{\ZR,\NR}{Z,N,\fvec{q}(\xi)}
$ 
involved in Eq.~\eqref{eq:pzrnr} to follow a Gaussian distribution as discussed in Sec.\ref{sec:res:analytic}. In the work of \cite{zdeb2017}, this convolution of the flux was also obtained from a random neck rupture mechanism.

In this section, we compare fission distributions obtained through the direct calculation of the projected mass and charge of the fragments with the ones obtained assuming analytic probability distributions for $\cprob{\ZR,\NR}{Z,N,\fvec{q}}$.
For each distribution, the mean $\mu$ is set to the average number of particles (neutrons or protons) in the fragments. Note that this slightly differs from our previous works \cite{regnier2016fission,regnier2019} where $\mu$ was chosen to be the closest integer to the average number of particles. 

The first analytical distribution we consider is the Gaussian distribution used in earlier works and also used to add a mass resolution in the fission fragment distribution before comparing with experimental data~\cite{jaffke2018hauser}. While the mean of this distribution is specified by the average number of particles, the standard deviation $\sigma$ is a free parameter. The contribution of a scission configuration is then
\begin{multline}
    \cprob{\ZR}{Z,N,\fvec{q}}
    =
        \frac{1}{2}
          \erf\left(
            \frac{\ZR+\frac{1}{2} - \mu(\fvec{q})}{\sqrt{2}\sigma}
          \right) \\
        - \frac{1}{2}\erf\left(
            \frac{\ZR-\frac{1}{2} - \mu(\fvec{q})}{\sqrt{2}\sigma}
          \right) .
\end{multline}

The hypothesis that the dispersion in charge number $\ZR$ in the fission fragments is minimal leads to the second distribution considered in this paper, the \textit{most local distribution} of mean $\mu$. The domain of this distribution is the set of positive integers. The only integers associated with a non-zero probability are $\lfloor\mu\rfloor$ and $\lfloor\mu\rfloor+1$, respectively having the probabilities $1-x$ and $x$, where $x=\mu-\lfloor\mu\rfloor$. Consequently, the contribution $\cprob{\ZR,\NR}{Z,N,\fvec{q}}$  reads
\begin{equation}
    \cprob{\ZR}{Z,N,\fvec{q}}
    =
      \begin{cases}
            1-x & \text{if $\Xfrag = \lfloor\mu(\fvec{q})\rfloor  $} \\
        \quad x & \text{if $\Xfrag = \lfloor\mu(\fvec{q})\rfloor+1$} \\
        \quad 0 & \text{otherwise}
      \end{cases}.\\
\end{equation}

If we model scission as a set of independent and identically distributed draws where each nucleons is put in the left or right fragment according to a coin tossing, the probability distribution follows a Binomial law
\begin{equation}
    \cprob{\ZR}{Z,N,\fvec{q}}
    =
      \binom{X}{\Xfrag} (\mu/X)^\Xfrag \left[1-(\mu/X)\right]^{X-\Xfrag}.
\end{equation}

Finally, the law of rare events suggests  that the more asymmetrical fission is, the better the binomial law can be approximated by the fourth and last distribution considered in this work, a Poisson distribution whose contribution at each point $\fvec{q}$ is
\begin{equation}
\cprob{\ZR}{Z,N,\fvec{q}}
    = \frac{\mu^{\Xfrag} e^{-\mu}}{\Xfrag !} .
\end{equation}

\begin{figure}[!ht] 
\begin{center} 
\includegraphics[width=\columnwidth]{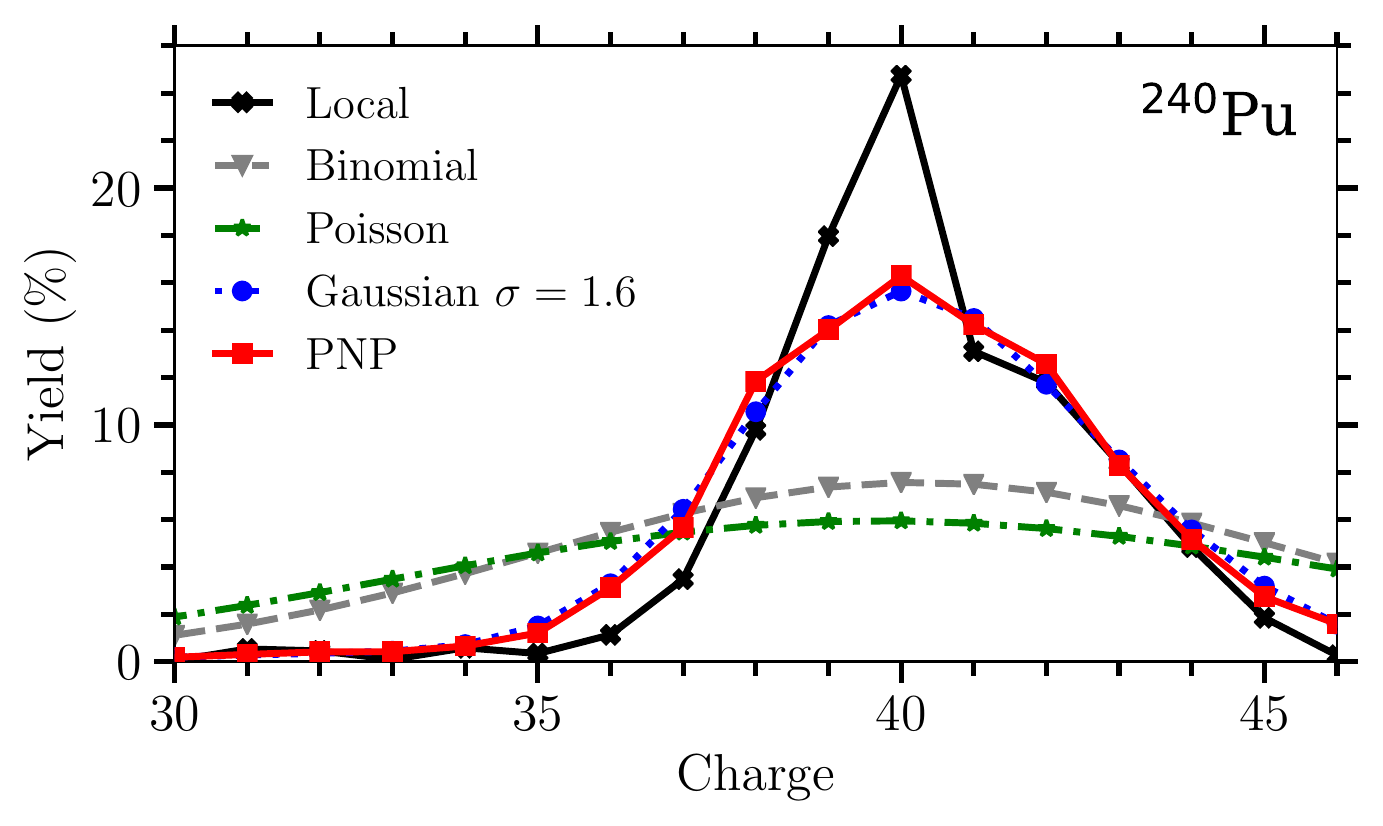}
\caption{\label{fig:YZ_pdf} Light peak of the charge distribution $Y(Z)$ in $^{240}$Pu as a function of $Z$ for different analytic forms for $\cprob{\ZR,\NR}{Z,N,\fvec{q}}$ as well as from PNP.
We only show the curve for the light fragment, since the whole distribution is completely symmetric with respect to $Z/2$; see text for additional details.}
\end{center} 
\end{figure}

The results are summarized in \figref{fig:YZ_pdf} for the charge yields; the mass yields are qualitatively similar. The Poisson and Binomial distributions are considerably too broad. In contrast, both the local discrete and Gaussian distribution do a decent job at reproducing the width of the charge distribution, although the local discrete distribution is too narrow leading to overestimating the height of the peaks. The fact that a Gaussian convolution reproduces quite well the PNP is consistent with the conclusion of Ref.~\cite{lacroix2020counting} although they were obtained in a different context where the projection subvolume is spherical.

\subsection{Evolution of Yields as Function of Frontier Definition}
\label{sec:res:qn}

One of the well-known limitations of adiabatic approaches to fission is the dependency of results on the definition of scission configurations \cite{davies1977calculation,younes2011,schunck2016microscopic}. This problem is especially relevant in self-consistent HFB calculations of PES: the large computational cost often leads to limiting the number of active collective variables -- two in our case, which leads to discontinuities in the PES \cite{dubray2012numerical}. Very often, scission configurations are in fact associated with such a major discontinuity. In \figref{fig:PES}, the line that separates the colored area from the white background at $q_{20} > 300$ b is the visual representation of such a discontinuity, where the expectation value $\qneck$ of the Gaussian neck operator drops from 3 -- 6 down to values smaller than 0.1. It is especially important to quantify the impact of the definition of the frontier on charge distributions as recent work suggest the odd-even staggering effect may only manifest itself at rather small values of the neck between the two prefragments \cite{verriere2020improvements}.

\begin{figure}[!ht] 
\begin{center} 
\includegraphics[width=\linewidth]{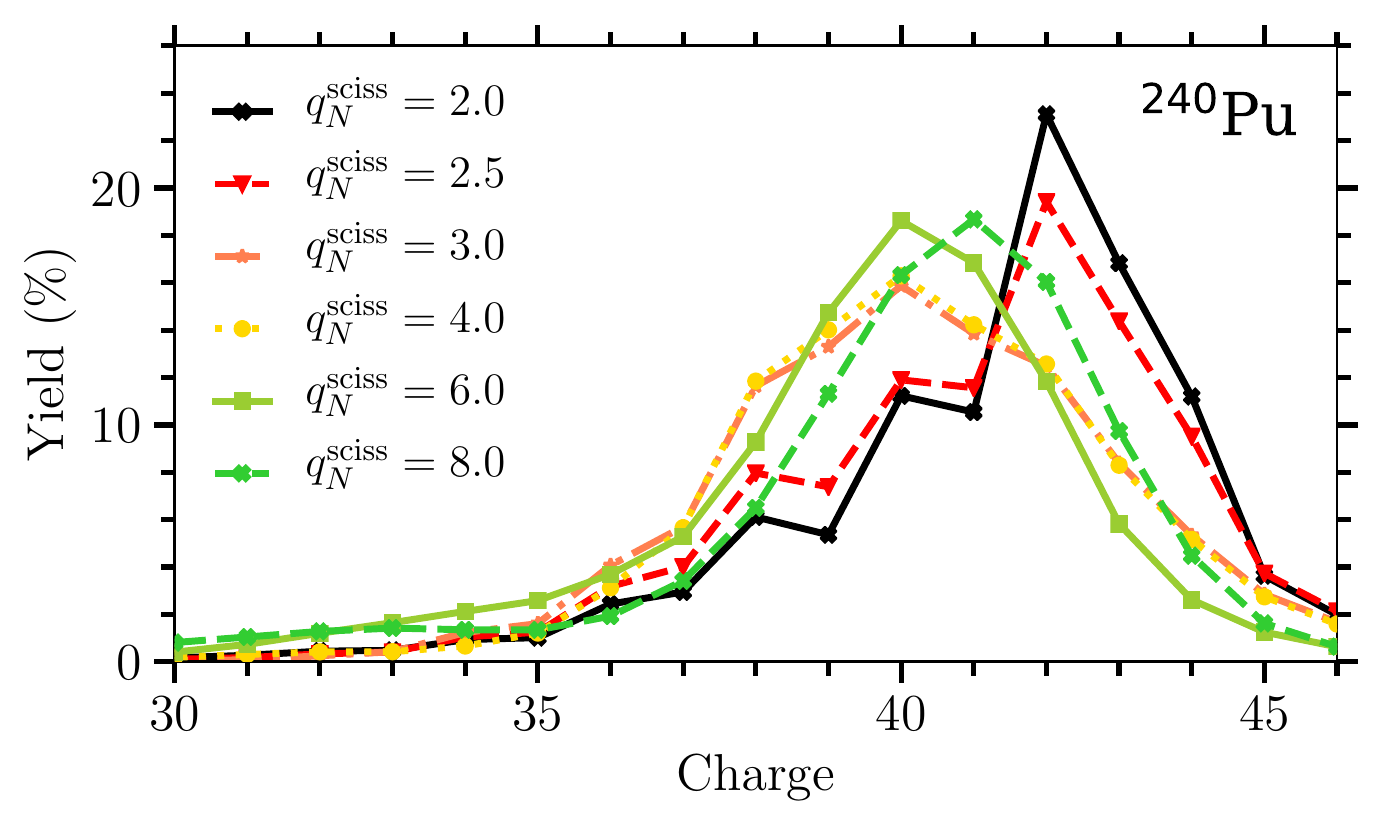}
\caption{\label{fig:YZ_qn_pdf} Light peak of the fragment charge distribution $Y(Z)$ for $^{240}$Pu at $E_n = 1.0$ MeV obtained with different values of the frontier definition $\qnecksciss$.} 
\end{center} 
\end{figure}

In \figref{fig:YZ_qn_pdf}, we thus show the evolution of the charge distribution in $^{240}$Pu as a function of the criterion $\qnecksciss$ used to define scission configurations, as explained in Sec.~\ref{sec:theo:proj:population}. As expected, there is no evidence of odd-even staggering for large values of $\qneck$. For $3.0 \leq \qneck \leq 8.0$, the peak does not really change very significantly. The first hints of odd-even staggering only appear at $\qneck \leq 2.5$. At the same time, there is a more pronounced shift of the peak toward larger values of $Z$. While this result is encouraging, we should point out that values of $\qneck < 3$ often require interpolating the PES through the discontinuity defining scission, which is non-physical. Better-resolved, continuous PES are required to verify the possibility of an odd-even staggering effects. This may be achieved for instance by increasing the number of collective degrees of freedom.

\begin{figure*}[!hbt] 
\begin{center} 
\includegraphics[width=\linewidth]{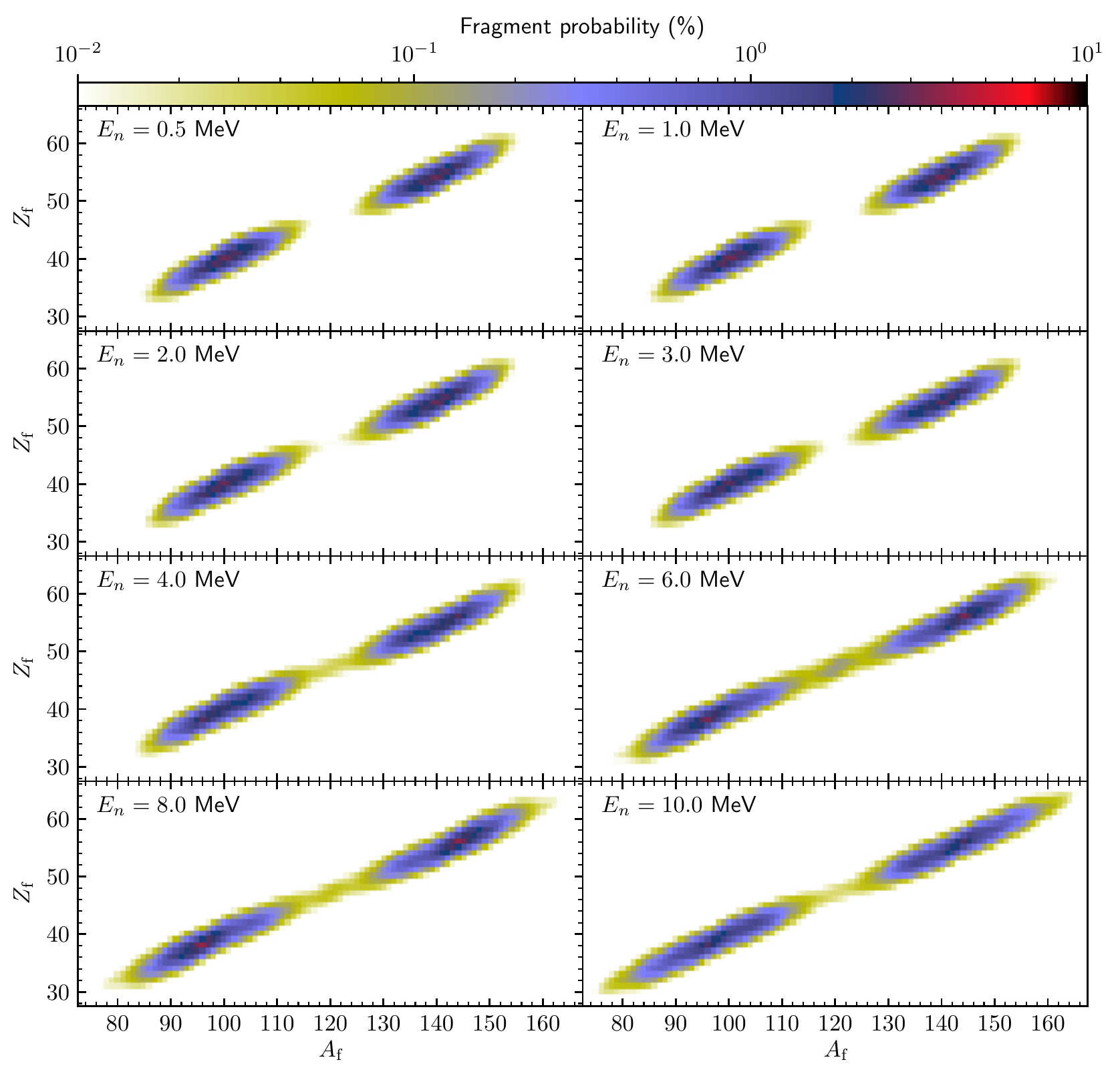}
\caption{\label{fig:YZA_Ex_pdf} Isotopic yields $Y(\Zfrag, \Afrag)$ for the reaction $^{239}$Pu(n,f) for different incoming neutron energy $E_n$ using PNP in the fragments. Symmetric fission becomes more probable with increasing excitation energy.} 
\end{center} 
\end{figure*}


\subsection{Evolution of Yields as Function of Excitation Energy}
\label{sec:res:En}

Within our approach, we can estimate the primary fission fragment mass and charge distributions $Y(A; E_n)$ and $Y(Z; E_n)$ for different values of the incoming neutron energy $E_n$. To this purpose, we make the approximation that all the excitation that the compound nucleus acquires after absorbing the neutron is of collective nature. This implies that varying incident neutron energies $E_n$ translate simply into varying mean energies of the collective wave packet; see Eq.~\eqref{eq:phi_init}. We are very aware that this approximation is rather strong and probably not entirely valid. A more accurate treatment of collective dynamics would require including quasiparticle excitations in the TDGCM formalism along the lines of~\cite{bernard2011}, or adopting a finite-temperature approach \cite{schunck2015b,zhao2019}. The former strategy faces considerable computational challenges, while the latter is not formally well-defined: the very concept of a GCM wave function made of a superposition of kets requires generalization at finite temperature; see, e.g., \cite{dietrich2010}. Pending such future developments and bearing in mind the limitations of our approach, we can still provide a useful reference point for future comparisons.

In \figref{fig:YZA_Ex_pdf}, we thus show the evolution of the full, two-dimensional isotopic fragment distribution as a function of incident neutron energy. As expected from statistical mechanics -- at ``infinite'' excitation energy, all fragmentations should become equiprobable -- the distribution $Y(Z,A; E_n)$ includes an ever larger set of fragmentations as the incident neutron energy increases from $E_n=0.5$ MeV to $E_n=10.0$ MeV. In particular, symmetric fission becomes more and more likely, which is especially visible for $E_n > 3.0$ MeV, and very asymmetric configurations with $A_{L} \approx 80$ get more and more populated. A more careful analysis of these maps show that the centroid of the distribution for the light fragment shifts toward lower $Z_{L}$ and lower $A_{L}$: at $E_n = 0.5$ MeV, it is located at $(Z_L,A_L) = (39.875, 100.364)$ and has moved to $(Z_L,A_L) = (37.987, 95.575)$ at $E_n = 10.0$ MeV. Similarly the most likely fission fragment changes from $^{100}$Zr ($Z_{L} = 40$, $N_{L} = 60$) at $E_n = 0.5$ MeV to $^{96}$Sr ($Z_{L} = 37.987$, $A_L = 95.575$) at $E_n = 10.0$ MeV.

\begin{figure}[ht] 
\begin{center} 
\includegraphics[width=\linewidth]{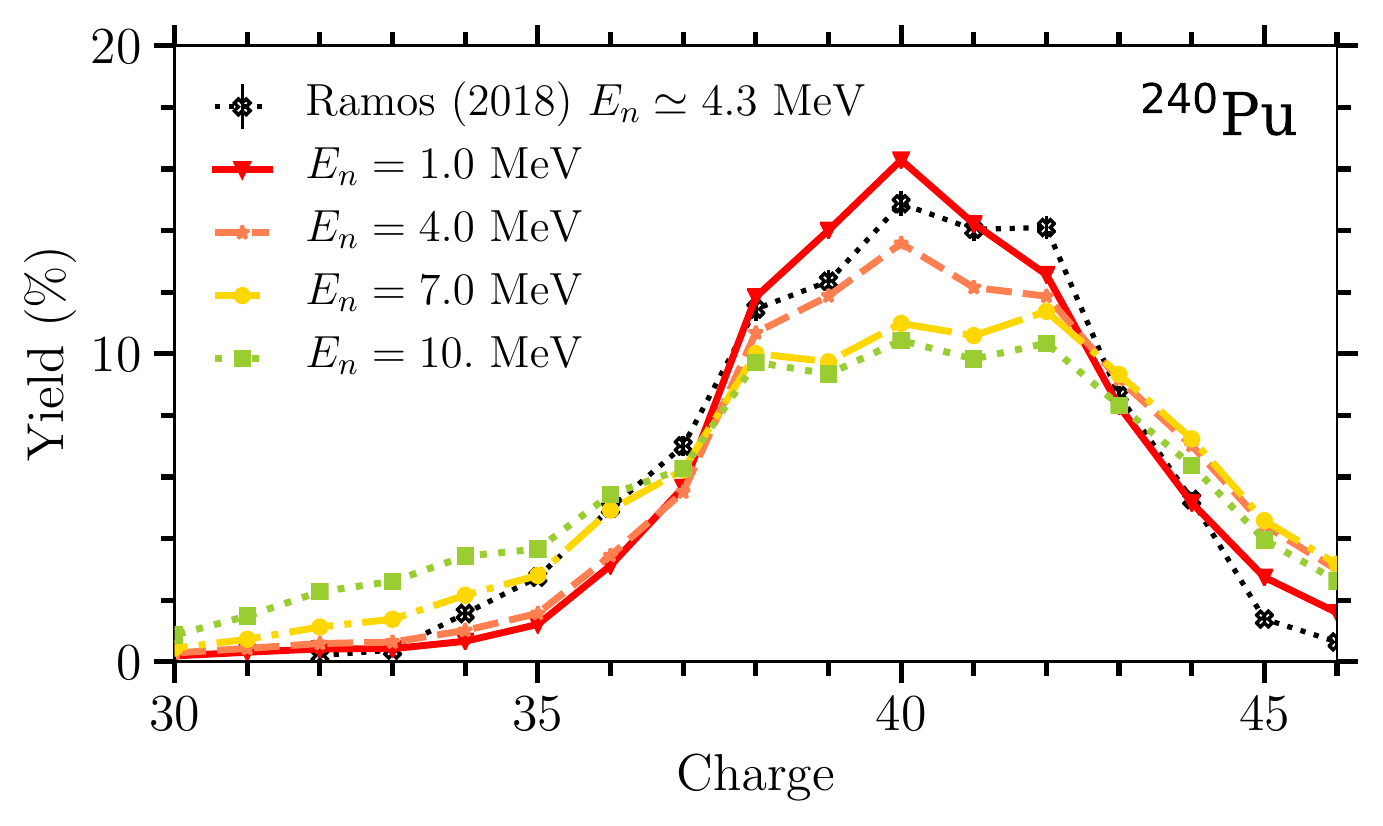}
\caption{\label{fig:YZ_Ex_pdf} 
Light peak of the fragment charge distribution $Y(Z)$ for $^{240}$Pu obtained with different values of initial energy of the compound nucleus. Results with PNP are compared to the experimental data from Ref.~\cite{ramosIsotopic2018}.}
\end{center} 
\end{figure}

To get a more quantitative feel for the variations of individual charge yields, we show in \figref{fig:YZ_Ex_pdf} the charge yields of $^{240}$Pu for various incident neutron energies in the range $E_n\in[0.5;10]$ MeV. These results are compared to the experimental data of D. Ramos \textit{et al.}~\cite{ramosIsotopic2018} obtained in inverse kinematics. This experiment leverages the transfer reaction $^{12}$C$(^{238}$U$,^{240}$Pu$)^{10}$Be to produce the compound Plutonium system with an average excitation energy of 10.7 MeV and a standard deviation of 3.0 MeV. This energy corresponds to an incident neutron energy of $E_n= E^* - S_n\simeq4.3$ MeV although differences on the resulting fission yields may arise from the difference of input reaction channel.
Once again our results show a broadening of the charge yields with the increase of the neutron energy. Overall, our predictions at $E_n=4$ MeV compare well with the experimental data of Ref.~\cite{ramos2018}. We still notice a strong overestimation of the predicted charge yields in the symmetric valley.

\begin{figure}[!ht] 
\begin{center} 
\includegraphics[width=\linewidth]{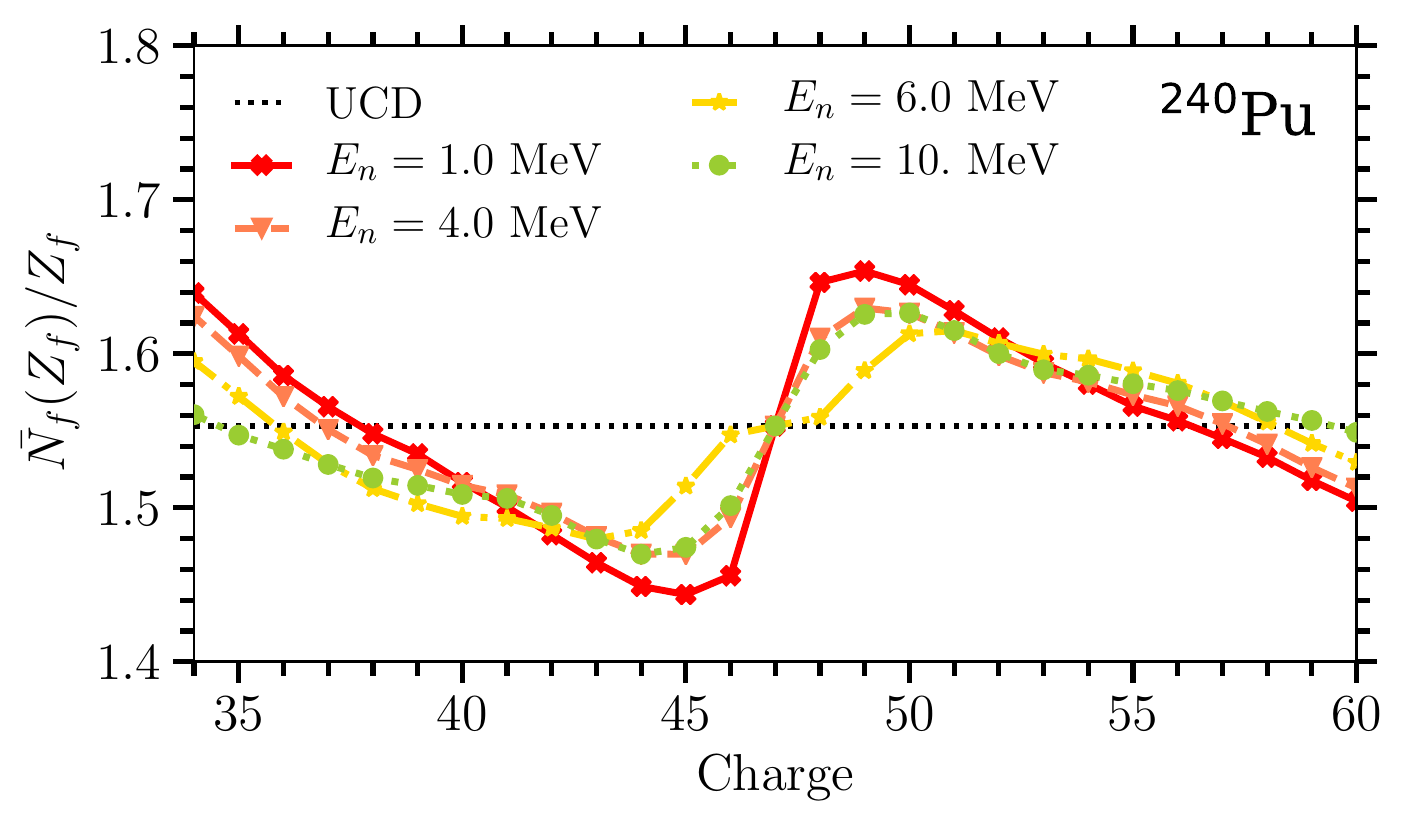}
\caption{\label{fig:ChargePol_Ex_pdf} Evolution of the neutron excess of the primary fission fragments for the reaction $^{239}$Pu(n,f) according to $E_n$. If the UCD were satisfied, all these quantities would be constant equal to 1.55.} 
\end{center} 
\end{figure}
Finally, \figref{fig:ChargePol_Ex_pdf} shows the evolution of the neutron excess of Eq.~\eqref{eq:n_excess} as a function of incident neutron energy. We see that the addition of the excitation energy to the system tends to flatten the structure of the neutron excess. These predictions are qualitatively consistent with the observed fission of $^{250}$Cf at high excitation energy ($E^*=43$ MeV)~\cite{caamanoCharacterization2015} as well as with the interpretation of shell effects being smoothed out as the energy increases.

\section{Conclusion}

In this paper, we reported the first microscopic calculation of fission fragment distribution within the TDGCM+GOA framework where the number of particles in fission fragment was extracted by direct particle number projection. While using Gaussian folding of particle number at scission is a good approximation of the exact PNP result, it still requires specifying the width. Even guided by consideration about pairing fluctuations or the size of the neck, this parameter remains somewhat arbitrary: PNP techniques allow eliminating it entirely and thereby obtaining more realistic distributions.

When comparing with experimental data for the two standard cases of $^{236}$U and $^{240}$Pu, we find that the agreement on both charge and mass yields is satisfactory, especially considering the rather large uncertainties on the primary fragment distributions, which translates itself into a large experimental mass resolution for the mass yields, and somewhat conflicting datasets for the charge yields. Our analysis of the two-dimensional isotopic yields recovers the excepted charge polarization already at the level of the primary yields, i.e., before any evaporation of neutrons. Many recent measurements seem to confirm the existence of an odd-even staggering in the charge distributions. Our calculations with small neck sizes for scission configurations seem to confirm this, although the resolution of the potential energy surface is not sufficient to draw firm conclusions. 

The evolution of the yields as a function of incident neutron energy, which in our framework translates into collective excitation energy of the fissioning nucleus, is compatible with what is expected from statistical mechanics and observed experimentally: the peaks of the distribution widen and symmetric fission increases. We note that the shell effects visible on the predicted charge polarization tends to vanish with increasing energy.

\begin{acknowledgments}
This work was supported in part by the NUCLEI SciDAC-4 collaboration DE-SC001822 and was performed under the auspices of the U.S.\ Department of Energy by Lawrence Livermore National Laboratory under Contract DE-AC52-07NA27344. Computing support came from the Lawrence Livermore National Laboratory (LLNL) Institutional Computing Grand Challenge program.
\end{acknowledgments}

\bibliography{biblio,zotero_output,books}

\end{document}